# Generically Automating Separation Logic by Functors, Homomorphisms, and Modules

**QIYUAN XU,** Nanyang Technological University, Singapore
**DAVID SANAN,** Singapore Institute of Technology, Singapore
**ZHE HOU,** Griffith University, Australia
**XIAOKUN LUAN,** Peking University, China
**CONRAD WATT,** Nanyang Technological University, Singapore
**YANG LIU,** Nanyang Technological University, Singapore

Foundational verification considers the functional correctness of programming languages with formalized semantics and uses proof assistants (e.g., Coq, Isabelle) to certify proofs. The need for verifying complex programs compels it to involve expressive Separation Logics (SLs) that exceed the scopes of well-studied automated proof theories, e.g., symbolic heap. Consequently, automation of SL in foundational verification relies heavily on ad-hoc heuristics that lack a systematic meta-theory and face scalability issues. To mitigate the gap, we propose a theory to specify SL predicates using abstract algebras including functors, homomorphisms, and modules over rings. Based on this theory, we develop a generic SL automation algorithm to reason about any data structures that can be characterized by these algebras. In addition, we also present algorithms for automatically instantiating the algebraic models to real data structures. The instantiation works compositionally, reusing the algebraic models of component structures and preserving their data abstractions. Case studies on formalized imperative semantics show our algorithm can instantiate the algebraic models automatically for a variety of complex data structures. Experimental results indicate the automatically instantiated reasoners from our generic theory show similar results to the state-of-the-art systems made of specifically crafted reasoning rules. The presented theories, proofs, and the verification framework are formalized in Isabelle/HOL.

## 1 Introduction

Foundational verification [13] involves verifying the functional correctness of concrete programs based on formal semantics using a proof assistant, in which both the inference and the proofs are certified. Therefore, foundational verification relies on a smaller trust base and produces more

Authors' Contact Information: Qiyuan Xu, Nanyang Technological University, Singapore, xu@qiyuan.me; David Sanan, Singapore Institute of Technology, Singapore, david.miguel@singaporetech.edu.sg; Zhe Hou, Griffith University, Brisbane, Australia, z.hou@griffith.edu.au; Xiaokun Luan, Peking University, Beijing, China, luanxiaokun@pku.edu.cn; Conrad Watt, Nanyang Technological University, Singapore, conrad.watt@ntu.edu.sg; Yang Liu, Nanyang Technological University, Singapore, and China-Singapore International Joint Research Institute (CSIJRI), Guangzhou, yangliu@ntu.edu.sg.

trustworthy results than other formal methods. These advantages have promoted the rapid development of the field in recent years [23, 47, 59, 63–65].

Foundational semantics often include complex and low-level resource models that involve aliasing or references [27, 39, 45, 74], in which case Separation Logic (SL) has shown to be an effective verification method [54]. In a typical workflow of an SL-based foundational verification, the process has the following steps: (1) extract SL entailments (i.e., implications between SL formulas) that imply program correctness (e.g., by a predicate transformer [31, 64]), (2) then extract pure proof obligations (e.g., first-order logic formulas) to entail the validity of the entailments, and finally, (3) the pure proof obligations are sent to Automated Theorem Provers (ATPs) [1, 59, 64] for solving.

Step 2 above is often the bottleneck of the automation. Despite an abundance of techniques for such extraction [2, 4, 20, 21, 29, 43, 44, 51, 52, 62, 67], their adoption in foundational verification faces three issues: (1) Existing techniques focus on a small fragment of SL, usually variants of *symbolic heap*, based on a simplified memory model having limited support to pointer arithmetic [4, 8, 34, 62, 68]. The use of SL in foundational verification can easily exceed this fragment by either unsupported connectives [37, 66], abstract models like partial commutative monoids [11, 18], or deeply formalized semantics involving real pointer arithmetic [49]. (2) Most techniques support only linked data structures like linked lists and trees [20, 43, 44, 57], whereas complex software heavily uses advanced data structures like arrays, dynamic arrays, and hash tables. (3) In foundational verification, scalability relies on abstractions that encapsulate complicated concrete details and provide abstract layers to ease the verification [24, 25, 40, 61]. Predicates are essential means to provide abstraction, whereas existing techniques rely on unfolding predicates, which can destroy such abstractions, blow up expressions, and finally cause scalability limitations.

The absence of solutions addressing these issues has motivated researchers to develop approaches for automating expressive SLs in foundational verification. Among them, RefinedC [64] represents a milestone. Based on Iris, it proposes a type-based specification language and a logic programming-based reasoning framework, providing automation that is on par with non-foundational methods.

While state-of-the-art tools like RefinedC have made significant progress, they still face limitations in providing automation support for user-defined data structures. In particular, automation support on user structures is limited to the generation of folding and unfolding rules. Further complex transformations depend on manually crafted typing rules that constitute ad-hoc reasoning procedures. Depending on the complexity of the structures involved, crafting the typing rules demands substantial expertise, coupled with extensive creative input and additional proving efforts.

We can see examples in non-trivial transformations such as subtyping $\text{List}(T) <: \text{List}(U)$, which is ubiquitous when containers have different element types; splitting and concatenating array slices, e.g., from $\text{int}[1..9]$ to $\text{int}[1..5] * \text{int}[6..9]$, which are essential in divide-and-conquer algorithms like Quicksort; or separating container abstractions along element components, e.g., from $\text{Array}(T*U)$ to $\text{Array}(T) * \text{Array}(U)$, useful when components are owned by different program modules. Considering these examples, there is an absence of a systematic theory that enables a generic mechanism supporting the automatic *generation* of reasoning rules for data structures and modalities.

In this work, borrowing the satisfaction operator from Hybrid Logic [5, 30], we present an interpretation of SL predicates from the perspective of data refinement (§4). This interpretation reveals properties of SL predicates (Table 1) that correspond to laws of abstract algebras, capable of modeling many data structures and modalities (§6). From these predicate properties, it is possible to automatically instantiate the necessary rules for the aforementioned transformations, providing an approach for the automatic generation of non-trivial[1] data structure reasoning rules, discharging users from providing manual specifications and additional proofs when proving certain properties over data structures. In particular, we can automate the process for any aggregated data structure, where users only need to manually specify and prove base datatypes, as is the case for integer datatypes in programming languages.

---
[1] meaning transformations that are not for folding or unfolding definitions

Table 1. Algebraic properties identified by the paper, over which our generic SL automation is built.

| Property | Description[*] |
| --- | --- |
| TF | denotes Functor of SL implications, useful for modeling data containers (e.g., List). It is the key to subtyping like $\mathrm{List}(T) <: \mathrm{List}(U)$, which allows a reasoning process to shift from the space of containers into the space of elements. |
| SH | Homomorphism of predicate operators over $*$, useful to extract components of elements in a container and to split container abstractions along the separation of element components. |
| SA | denotes Associativity of module-like predicate operators, useful to model concatenation of paths to resources, such as a file path or the path to a member field in a nested record. |
| SD | Scalar Distributivity of module-like predicates is useful for modeling split and concatenation of data structure slices, like array slices. |
| IE | denotes Identity Elements used to specify the abstractions representing empty. |
| Tr | denotes Transitivity and equivalences between abstractions. |

[*] Predicate and refinement relations are interchangeably used as they are the *same* thing in our system.

The instantiated transformation rules are powerful enough to constitute the core of an SL reasoner for an generic imperative heap language, as demonstrated in §10. Specifically, this reasoner first extracts SL entailments that entail the functional correctness of a given program, using a typical *wp* or *sp* transformer (§7.2). Then, it reduces the decision problems of the entailments to transformations between predicates (i.e., subtypings in terms of RefinedC) using bi-abduction (§8.2.1, §8.2.2). Finally, it applies the automatically instantiated transformation rules to extract verification conditions that entail the transformations' validity (§6). The verification conditions are sent to SMT solvers or handed to users. Consequently, we present a *generic* SL reasoner over the abstractions of the algebras listed in Table 1. It generally supports any data structure or modalities that satisfy (even some of) the algebras.

We evaluate this reasoner through 10 widely-used data structures and 592 lines of programs in formalized semantics in Isabelle/HOL. In most cases, our reasoner is able to prove the properties with less human intervention compared with state-of-the-art foundational verification tools.

In summary, the main contributions of this paper are summarized as follows:

(1) A set of algebraic properties that captures general transformations between refinements of data structures. Transformation rules for specific data structures are instantiated automatically once the properties of the structures are proven.
(2) A *generic* SL reasoner that: a) Uses rule instances of the algebraic properties for automatic inference of SL entailments in a compositional manner minimizing the need for unfolding; b) Allows automatic proving of the algebraic properties of predicates, minimizing even further manual proving from users.
(3) The Isabelle/HOL formalization for the algebraic properties and the SL reasoner.
(4) The evaluation of our reasoner and its formalization on a battery of examples involving 10 data structures in a total of 592 lines of 8 programs.

## 2 Motivating Example

Let us consider verifying a two-line program { alloc_data (1); free_data (1) }, which necessarily requires a non-trivial transformation, or in RefinedC, a manually proven typing rule.

```
struct list { void* data; list* next; };          void free_data (list* l) {
void* safe_alloc (size_t s) {                         if (l) { free (l->data); free_data (l->next); }
    void* ret = alloc (s);                        }
    if (ret) return ret; else abort ();           void verify_this (list* l) {
}                                                     alloc_data (l);
void alloc_data (list* l) {                           free_data (l);
    if (l) { l->data = safe_alloc(42); alloc_data (l->next); }   }
}
```

Before delving into the example, we first introduce a refinement-based assertion language simplified from recent refinement-type approaches [59, 64, 65] — we use *SL predicate* to represent *data refinement*. We interpret a predicate $T$ as *a refinement relation* that relates concrete constructs (like memory heaps) to abstractions. Specifically, $T(x)$ defines the set of concrete constructs that refine abstraction $x$. The notion of refinement type in the recent works [59, 64, 65] corresponds to SL predicates in our theory. To emphasize this correspondence and to be intuitive, we introduce the **notation** $x \mathbin{\text{\textnormal{\textbf{\textsf{\textasciitilde}}}}} T$ to abbreviate predicate application $T(x)$, i.e., $x \mathbin{\text{\textnormal{\textbf{\textsf{\textasciitilde}}}}} T \triangleq T(x)$.

Returning to the above example, we fix a predicate $T$ to represent the refinement relation of the data in the linked list. The refinement of the linked list itself is then specified by a predicate operator $\text{List}_a : \text{Predicate} \to \text{Predicate}$, where $a$ denotes the address of the list. Predicate application $l \mathbin{\text{\textnormal{\textbf{\textsf{\textasciitilde}}}}} \text{List}_a(T)$ relates the memory heap of a concrete linked list to a sequence $l$, where the $i^{\text{th}}$ element of $l$ is the abstraction of the data in the $i^{\text{th}}$ linked node. To account for potentially null data pointers, we use the sum operator $(T + \text{Null})$ to represent a data entry that may be null[2]. There is an implication $x \mathbin{\text{\textnormal{\textbf{\textsf{\textasciitilde}}}}} T \longrightarrow (\text{inj}_1 \, x) \mathbin{\text{\textnormal{\textbf{\textsf{\textasciitilde}}}}} (T + \text{Null})$ that corresponds to a subtyping: the type of non-null pointers is a subtype of the nullable pointer type.

The safe_alloc always returns a non-null pointer, while a null pointer is a valid argument of free. Therefore, we stipulate the postcondition of routine alloc_data to be $l \mathbin{\text{\textnormal{\textbf{\textsf{\textasciitilde}}}}} \text{List}_a(T)$, representing a list of non-null pointers. For routine free_data, we stipulate its precondition to be $l' \mathbin{\text{\textnormal{\textbf{\textsf{\textasciitilde}}}}} \text{List}_a(T + \text{Null})$, representing a list of nullable pointers. Given the specifications, if we want to verify { alloc_data (l); free_data (l)) }, we must prove an SL entailment $l \mathbin{\text{\textnormal{\textbf{\textsf{\textasciitilde}}}}} \text{List}_a(T) \longrightarrow l' \mathbin{\text{\textnormal{\textbf{\textsf{\textasciitilde}}}}} \text{List}_a(T + \text{Null})$, where $l' = \text{List.map}(\text{inj}_1)(l)$. This entailment corresponds to a subtyping: a list of non-null pointers is a subtype of a list of nullable pointers. Importantly, while this subtyping seems intuitive, it cannot be derived automatically by state-of-the-art tools such as RefinedC.

Taking RefinedC as an example, non-null pointers are specified by type $\&_{\text{own}}$ while nullable pointers are by optional($\&_{\text{own}}$, null). The system provides subtyping $\&_{\text{own}} <: \text{optional}(\&_{\text{own}}, \text{null})$. However, it cannot derive subtyping $\text{list}(\&_{\text{own}}) <: \text{list}(\text{optional}(\&_{\text{own}}, \text{null}))$. This limitation stems from RefinedC's restricted automation support for user-defined types. While RefinedC automatically derives (un)folding rules for user-defined types, it does not generate subtyping rules for them. This gap necessitates manual intervention to prove the subtyping rule of user-defined types. This task requires extensive expertise in understanding the internal implementation of RefinedC and in proving SL lemmas using Iris, therefore presenting a significant challenge for non-experts.

In our system, a mechanism is provided for deriving subtyping rules of user-defined predicate operators automatically. Continuing the example, the subtyping rule of List is represented as

$$\dfrac{\forall e \in \text{set}(l).\ e \mathbin{\text{\textnormal{\textbf{\textsf{\textasciitilde}}}}} T \longrightarrow f(e) \mathbin{\text{\textnormal{\textbf{\textsf{\textasciitilde}}}}} U}{l \mathbin{\text{\textnormal{\textbf{\textsf{\textasciitilde}}}}} \text{List}_a(T) \longrightarrow \text{map}(f)(l) \mathbin{\text{\textnormal{\textbf{\textsf{\textasciitilde}}}}} \text{List}_a(U)} \quad \begin{array}{l} \text{map}(f)([l_1, \cdots, l_n]) \triangleq [f(l_1), \cdots, f(l_n)] \text{ is the mapper of lists;} \\ \text{set}([l_1, \cdots, l_n]) \triangleq \{l_1, \cdots, l_n\} \text{ gives the set of elements in a list} \end{array}$$

The premise assumes that when the refinement relation of an element changes from $T$ to $U$, its abstraction changes from $e$ to $f(e)$. The conclusion then shows that the refinement relation of the entire list can change from $\text{List}_a(T)$ to $\text{List}_a(U)$, and its abstraction changes accordingly from $l$ to $\text{map}(f)(l)$. In essence, this rule specifies how to transform the refinement of a linked list based on the transformation of the refinements of its elements.

---

[2]The sum operator is generally defined as $(\text{inj}_1 \, x) \mathbin{\text{\textnormal{\textbf{\textsf{\textasciitilde}}}}} (T_1 + T_2) \triangleq x \mathbin{\text{\textnormal{\textbf{\textsf{\textasciitilde}}}}} T_1$ and $(\text{inj}_2 \, y) \mathbin{\text{\textnormal{\textbf{\textsf{\textasciitilde}}}}} (T_1 + T_2) \triangleq y \mathbin{\text{\textnormal{\textbf{\textsf{\textasciitilde}}}}} T_2$ where $\text{inj}_1$ is the left injection while $\text{inj}_2$ is the right.

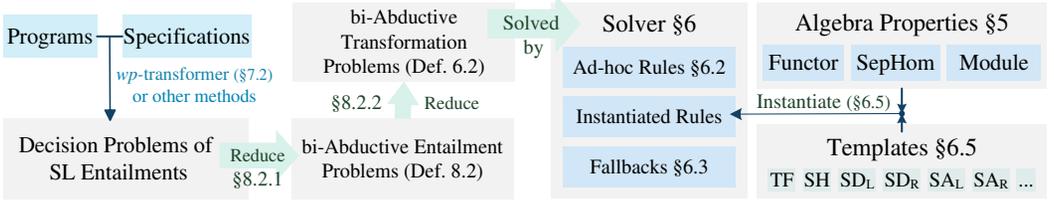

Fig. 1. Overall workflow of our Separation Logic reasoner.

The target proof goal $l \mathbin{\S} \text{List}_a(T) \longrightarrow l' \mathbin{\S} \text{List}_a(T + \text{Null})$ is then derived from this subtyping rule, the known fact $\forall e.\ (e \mathbin{\S} T \longrightarrow (\text{inj}_1\ e) \mathbin{\S} (T + \text{Null}))$, and a proof obligation $l' = \text{map}(\text{inj}_1)(l)$.

This subtyping rule reflects the covariant functor property of List over the SL implication ($\longrightarrow$). Indeed, the rule is instantiated from an instance Functor(List, map, set) of a generally defined algebraic property Functor($F, m, d$) specifying that predicate operator $F$ is a functor over ($\longrightarrow$),

$$\text{Functor}(F, m, d) \triangleq \text{ for any } f, x \text{ and predicate } T, T', \qquad\qquad\qquad \textbf{(TF)}$$
$$x \mathbin{\S} F(T) \longrightarrow m(f)(x) \mathbin{\S} F(T') \text{ holds if } \forall e \in d(x).\ e \mathbin{\S} T \longrightarrow f(e) \mathbin{\S} T' \text{ holds.}$$

Assuming $F$ represents some data container, then $m$ represents the mapper of (the abstraction of) the container, e.g., list mapper; $d(x)$ gives the element domain of (the abstraction of) a container instance $x$. This property then provides a generic mechanism for generating subtyping rules for any predicate operators satisfying the property.

If a predicate operator $F$ is defined as a composition of other functors, say $F \triangleq G_1 \circ \cdots \circ G_n$, our reasoner is able to derive $F$'s functor property automatically by Functor composition — Functors $G_1, G_2$ yield $G_1 \circ G_2$ as a functor.

$$\frac{\text{Functor}(G_1, m_1, d_1) \qquad \text{Functor}(G_2, m_2, d_2)}{\text{Functor}(G_1 \circ G_2, m_1 \circ m_2, d_1 \ggg d_2)} \text{ (Functor Composition),}$$

where $(d_1 \ggg d_2) \triangleq (\lambda x.\ \bigcup_{e \in d_1(x)} d_2(e))$ is the monadic bind of sets. This composition principle allows us to prove the algebraic properties of a composite predicate operator without unfolding its component operators. For recursively defined predicate operators, our reasoner applies an induction tactic, as elaborated later in §9.

## 3 Overview

This work aims to increase the automation of program verification by adding automated support to user-defined data structures. Such support comes from two different aspects: automated rule generation and automated SL entailment reasoning by means of the automated generated rules.

By identifying algebraic structures encoding refinement transformation scenarios between predicates, presented in §5, it is possible to design a generic reasoner that has the knowledge of the scenarios and can automate refinement of any data structure satisfying the algebraic axioms in the scenarios. This approach leads to a systematic method for automating our SL logic (§4).

To facilitate this, the reasoner centers around a designed family of problems called *bi-abductive Transformation Problems (bi-TPs)*. These problems are formulated to represent refinement transformations of the form $x \mathbin{\S} T \longrightarrow f(x) \mathbin{\S} U$, allowing direct application of the algebraic properties.

Illustrated in Fig. 1, our reasoning process unfolds in two stages: (1) introduced in §7 and §8 a SL reasoner reduces program verification problems to bi-TPs; (2) introduced in §6 a TP-Solver applies rules automatically generated based on algebraic properties of predicates to solve the bi-TPs. The algorithms for proving the algebraic properties of a given predicate are provided in §9.

## 4 A Separation Logic with a Perspective of Data Refinement

The theory of our Separation Logic (SL) automation algorithm is based on interpreting SL predicates as data refinement relations. Before presenting the algorithm, we have to first formalize how this interpretation is established on an SL semantics, and we also formalize this SL semantics. For brevity, this SL formalization is simplified while the complete version is left to Appendix A.

The assertion language of our SL is parameterized by a finite set $\mathbf{P}$ of SL predicates, and a first-order logic *FOL* with equality. Let $w, x, y, z, t$ range over terms in *FOL*, $\alpha, \beta$ over variables in *FOL*, and $T, U$ over SL predicates $\mathbf{P}$. The assertion language $\mathbf{F}$ of our SL includes all standard connectives plus a satisfaction operator ($\Vdash$) borrowed from Hybrid Logic [5, 30] (originally denoted by @).

$$\mathbf{F} \ni \phi, \psi ::= \top \mid \bot \mid \mathsf{emp} \mid T(x) \mid \neg \phi \mid \phi * \psi \mid \phi \wedge \psi \mid \phi \vee \psi \mid \phi \twoheadrightarrow \psi \mid \phi \rightarrow \psi \mid \exists \alpha. \phi \mid \forall \alpha. \phi \mid t \Vdash \phi$$
$$\mid \text{any other formula in } FOL, \text{ e.g., } x = y.$$

Ranged over by $F, G$, the set $\mathbf{P}^*$ of SL *predicate operators* is defined as an inductive set consisting of all predicates in $\mathbf{P}$, all maps from $\mathbf{P}$ to $\mathbf{P}^*$, and all maps from *FOL* terms to $\mathbf{P}^*$. This notion of predicate operator will be used later in formalizing algebraic properties. The logic is first-order and does not support quantifying over predicates. Predicate operators are defined in the meta-logic.

The semantics of an SL is conventionally defined upon a partial algebra known as *Separation Algebra* (SA). There are various ways in the literature to define the notion of SA. We follow a widely accepted one [11] that defines an SA as a Partial Commutative Monoid (PCM), written $(S, \bullet, \epsilon)$ for a carrier set $S$, a partial binary operation $\bullet$ over $S$, and an identity element $\epsilon \in S$.

Fix a domain of discourse $O$ and an interpretation function $[\![-]\!]$ from *FOL* terms to $O$. Fix a PCM $\mathcal{A} = (S, \bullet, \epsilon)$ such that $O \subseteq S$. Elements in $S$ are called *worlds* and ranged over by $w$.

The semantics of formulas in $\mathbf{F}$ is defined by forcing relation ($\models$), a binary relation between $S$ and $\mathbf{F}$. Note that ($\models$) should be parameterized by the interpretation function that maps every predicate to a subset of $O$, but we omit it here. For $w \in S$, and $\phi, \psi \in \mathbf{F}$,

$w \models \phi * \psi$ holds iff there exists $w_1, w_2$ such that $w = w_1 \bullet w_2$ and both $w_1 \models \phi$, $w_2 \models \psi$ hold.
$w \models \phi \twoheadrightarrow \psi$ holds iff for any $w'$ such that $w' \models \phi$ holds and $w' \bullet w$ is defined, $w' \bullet w \models \psi$ holds.
$w \models (t \Vdash \phi)$ holds for any term $t$ iff the world represented by term $t$ satisfies formula $\phi$, i.e. $[\![t]\!] \models \phi$ holds. Essentially, ($\Vdash$) internalizes the forcing relation ($\models$) into the object logic.

The definition of ($\models$) on predicate application and other connectives is standard and omitted here.

*Definition 4.1 (Validity).* We say an SL formula $\phi$ holds, iff $\forall w \in S.\ w \models \phi$ holds.

*Definition 4.2 (Type notation).* $x \mathbin{\text{\textfrac{9}{}}} T \triangleq T(x)$, for any term $x$ and predicate $T$.

*Definition 4.3 (Data Refinement implied by SL predicates).* Iff $w \models x \mathbin{\text{\textfrac{9}{}}} T$ holds, we say $w$ refines $x$, or equivalently, $x$ is an abstraction of $w$, w.r.t. refinement relation $\hat{T} \triangleq \{(w, x) \mid (w \models T(x))\}$.

$x \mathbin{\text{\textfrac{9}{}}} T$ relates a set of concrete constructs to one specific abstract object $x$. In order to relate concrete constructs to a set of abstract objects, one can use ($\exists$), e.g., $\exists a.\, a \mathbin{\text{\textfrac{9}{}}} T \wedge a \in A$ which specifies concrete constructs that refine one of the abstract objects in set $A$.

Turning back to ($\Vdash$), there are two reasons to introduce ($\Vdash$). First, it allows us to specify stepwise refinement, $x \mathbin{\text{\textfrac{9}{}}} (U; T) \triangleq (\exists y.\, y \mathbin{\text{\textfrac{9}{}}} U \wedge (y \Vdash x \mathbin{\text{\textfrac{9}{}}} T))$. Second, it allows us to use and express a predicate as a refinement relation $\hat{T}(w, x) \triangleq (w \Vdash T(x))$ within the logic. This explicit expression of refinement relation is essential. For example, assuming $a \mapsto v$ is an assertion specifying a singleton heap which has value $v$ at address $a$, the predicate $x \mathbin{\text{\textfrac{9}{}}} \mathrm{Ref}_a(T) \triangleq \exists v.\, (a \mapsto v) \wedge (v \Vdash x \mathbin{\text{\textfrac{9}{}}} T)$ specifies a memory object at address $a$ that has a value refining $x$ w.r.t. $T$.

We require the PCM $\mathcal{A}$ of our SL to encompass *all elements* in the domain of discourse $O$ and all concrete constructs including memory heaps and *program values*. This allows us to write our refinement relations as regular SL predicates to abstract any constructs we want. Examples include the definition of stepwise refinement $T; U$ above (which requires $y$ in the PCM), and the definition of Ref which requires $v$ in the PCM.

We use SL predicates to build refinements because (1) we want to use $(*)$ to separately and compositionally specify the refinement of each component, e.g. $\text{Ref}_a(T * U)$ for a reference containing two components; (2) instead of introducing a different relational separating conjunction to express the $T * U$ in $\text{Ref}_a(T * U)$, we want every connective and operator to stay in a uniform category so that we could use a uniform automation mechanism for all of them, e.g., the same mechanism can be used to reason about both $\text{Ref}_a(T * U)$ and $(T * U)$.

This big PCM $\mathcal{A}$ may seem cumbersome, and its group operation is hard to define when elements belonging to different sorts are involved. In implementations, we adopt a many-sorted variant SL through a shallow embedding into typeclasses provided by the underlying proof assistants. This allows us to define the group operation individually for each sort of PCM. This many-sorted version is left to Appendix B while the paper's theoretical discussion only considers the single monolithic PCM, for the sake of simplicity.

Returning to unifying SL predicates and refinement relations, we axiomatically declare a predicate Id to represent identity refinement. Semantically, $w \models (x \,\fatsemi\, \text{Id})$ holds iff $w = x$.

By means of Id, we can represent any relation $R$ as an SL predicate $\hat{R}(x) \triangleq (\exists w.\ w \,\fatsemi\, \text{Id} \land R(w, x))$. Together with $\hat{T}$ that represents an SL predicate as a refinement relation, it unifies SL predicates and refinement relations. Indeed, $\hat{T} = T$ and $\hat{R} = R$; this hat operator is a bijection between SL predicates and refinement relations. It unifies SL predicates and refinement relations, thus justifying

*Notation. Every predicate and its corresponding refinement relation are denoted by the same symbol.*

By interpreting SL predicates as refinement relations, implications between SL predicate applications have a meaning of *transformations* between refinement relations: Formula $x \,\fatsemi\, T \longrightarrow y \,\fatsemi\, U$ represents that any semantic construct that refines $x$ w.r.t. $T$ also refines $y$ w.r.t. $U$.

## 5 Algebras of Refinement Transformations

Our methodology is based on the assumption that typical SL entailments in real program verification are synthesizable from finite algebraic scenarios of refinement transformations.

This algebraic approach also offers an additional advantage: it allows us to abstract away concrete details, reveal common properties among diverse constructs, and then automate them generically in one unified mechanism. As an example, which we will explore later, fractional permission has the same model, module-like structure, as array slices. Without an algebraic approach, it is hard to recognize that fractional permissions and array slices can be automated by one mechanism.

While our implementation explores additional algebraic structures, we present three key algebraic structures in this section. Their axioms are defined in Fig. 2. Parameterized over the axioms of the algebras, a generic SL reasoner is presented in the subsequent sections of the paper.

### 5.1 Transformation Functor (TF)

Functor (from Fig. 2) captures the essence of covariant subtyping. Assume predicate operator $F(\cdot)$ represents a data container such as $\text{List}(\cdot)$. Property $\text{Functor}(F, m, d)$ specifies how to transform the refinement of the container's elements. Specifically, consider a data container instance represented by $x \,\fatsemi\, F(T)$. Predicate $T$ represents the refinement relation of the container's elements. The domain of the elements in the instance is represented by $d(x)$. Besides, if a function $f$ can transform the abstractions of the elements from refinement relation $T$ to refinement relation $U$, then the abstraction of the container instance can be transformed from $x \,\fatsemi\, F(T)$ to $m(f)(x) \,\fatsemi\, F(U)$.

$$\text{Functor}(F, m, d) \triangleq \text{for any } T, U, f, \text{ and } x \in \text{dom}(m(f)), \quad \text{(TF)}$$
$$x \mathbin{\text{\S}} F(T) \longrightarrow m(f)(x) \mathbin{\text{\S}} F(U) \text{ holds if } \forall e \in d(x).\, e \mathbin{\text{\S}} T \longrightarrow f(e) \mathbin{\text{\S}} U \text{ holds.}$$

$$\text{SepHom}(F, s, z) \triangleq \text{for any } T, U, \text{ any } x \in \text{dom}(z) \text{ and } y \in \text{dom}(s), \quad \text{(SH)}$$
$$x \mathbin{\text{\S}} (F(T) * F(U)) \longrightarrow z(x) \mathbin{\text{\S}} F(T*U) \text{ and } y \mathbin{\text{\S}} F(T*U) \longrightarrow s(y) \mathbin{\text{\S}} (F(T) * F(U)) \text{ hold.}$$

*The following properties are parameterized by a ring-like algebra $(\mathcal{S}, +, \cdot)$ called scalar algebra.*

$$\text{Assoc}(F, g, h) \triangleq \text{for any } T, \text{ any } n, m \in \mathcal{S}, x \in \text{dom}(g_{n,m}) \text{ and } y \in \text{dom}(h_{n,m}) \quad \text{(SA)}$$
$$x \mathbin{\text{\S}} F_n(F_m(T)) \longrightarrow g_{n,m}(x) \mathbin{\text{\S}} F_{n\cdot m}(T) \text{ and } y \mathbin{\text{\S}} F_{n\cdot m}(T) \longrightarrow h_{n,m}(y) \mathbin{\text{\S}} F_n(F_m(T)) \text{ hold.}$$

$$\text{Dist}(F, s, z) \triangleq \text{for any } T, \text{ any } n, m \in \mathcal{S}, x \in \text{dom}(s_{n,m}) \text{ and } y \in \text{dom}(z_{n,m}), \quad \text{(SD)}$$
$$x \mathbin{\text{\S}} F_{n+m}(T) \longrightarrow s_{n,m}(x) \mathbin{\text{\S}} (F_n(T)*F_m(T)) \text{ and } y \mathbin{\text{\S}} (F_n(T)*F_m(T)) \longrightarrow z_{n,m}(y) \mathbin{\text{\S}} F_{n+m}(T) \text{ hold.}$$

$$\text{SUnit}(F, g, h) \triangleq \text{for any } T, \text{ identity scalar } \epsilon \in \mathcal{S}, \text{ any } x \in \text{dom}(g_\epsilon) \text{ and } y \in \text{dom}(h_\epsilon), \quad \text{(S1)}$$
$$x \mathbin{\text{\S}} F_\epsilon(T) \longrightarrow g_\epsilon(x) \mathbin{\text{\S}} T \text{ and } y \mathbin{\text{\S}} T \longrightarrow h_\epsilon(y) \mathbin{\text{\S}} F_\epsilon(T) \text{ hold.}$$

$$\text{SZero}(F, D) \triangleq x \mathbin{\text{\S}} F_0(T) \longleftrightarrow \text{emp holds for any } T, \text{ any zero scalar } 0 \in \mathcal{S} \text{ and } x \in D. \quad \text{(S0)}$$

Fig. 2. Algebras of common refinement transformations. *Notation* $x \mathbin{\text{\S}} T \triangleq T(x)$ *denotes predicate application.*

In an expressive refinement-based assertion language, predicates and predicate operators (types and type operators) are usually hierarchically combined, forming a structure reminiscent of nested "Matryoshka dolls". Functor enables a reasoning process to systematically "unpack" these nested structures layer by layer, reducing reasoning problems about containers to problems about their elements, recursively until reaching atoms. Continuing the example in §2, in order to prove the transformation (subtyping) from $l \mathbin{\text{\S}} \text{List}_a(T)$ to $l' \mathbin{\text{\S}} \text{List}_a(U)$, Functor indicates that it suffices to show that any element $e \mathbin{\text{\S}} T$ in the list can transform to $e' \mathbin{\text{\S}} U$ for some $e, e'$ constrained by $l, l'$.

Additionally, we call this property Functor because of its categorical correspondence. Given a category $C$ where its objects are all SL predicates and its morphisms between two objects $T, U$ are partial functions $f$ such that $\forall x \in \text{dom}(f).\, x \mathbin{\text{\S}} T \longrightarrow f(x) \mathbin{\text{\S}} U$, a Functor$(F, m, d)$ describes a functor in $C$, where predicate operator $F$ is its object function and $m$ is its morphism function ($d$ plays a minor role restricting the domains of morphisms).

Regarding generality, Functor is ubiquitous in data structures. In particular, most data containers are Functor instances. Furthermore, many modalities and connectives are also Functors, such as the operator Ref. $x \mathbin{\text{\S}} \text{Ref}_a(T)$ specifies a memory object at address $a$ has a value refining $x$ w.r.t $T$. It satisfies Functor$(\text{Ref}_a, \lambda x.\{x\}, \lambda f.f)$. Another example is permission modality $x \mathbin{\text{\S}} n \oplus T$, which claims ownership of an $n$ fraction of an object $x \mathbin{\text{\S}} T$, for a fraction $0 \leq n \leq 1$. When $n = 1$, it represents total permission that permits read and write; when $0 < n < 1$, it permits read-only access; when $n = 0$, $x \mathbin{\text{\S}} 0 \oplus T$ equals empty. It has a property Functor$(n \oplus, \lambda x.\{x\}, \lambda f.f)$. As one more example, consider the Later modality $\triangleright$ seen in many impredicative SLs [1, 37]. Its predicate version $x \mathbin{\text{\S}} \triangleright T \triangleq \triangleright(x \mathbin{\text{\S}} T)$ satisfies Functor$(\triangleright, \lambda x.\{x\}, \lambda f.f)$ because of the so-called mono rule $(\phi \longrightarrow \psi) \longrightarrow (\triangleright \phi \longrightarrow \triangleright \psi)$. Indeed, any modality that has such a mono rule is a Functor.

### 5.2 Separating Homomorphism (SH)

A separating homomorphism, given by SepHom$(F, s, z)$, specifies that the predicate operator $F$ is homomorphic over the predicate separating conjunction $(*)$. Recall the predicate separating conjunction $(x, y) \mathbin{\text{\S}} (T*U) \triangleq (x \mathbin{\text{\S}} T) * (y \mathbin{\text{\S}} U)$. Assuming that $x \mathbin{\text{\S}} F(T*U)$ represents a data container, then $T * U$ represents that every element in the container contains two components abstracted respectively by $T$ and $U$. SepHom$(F, s, z)$ allows us to split the abstraction of the container along the separation between the two element components and also merge the two divided parts back. The split operation $s$ transforms $x \mathbin{\text{\S}} F(T * U)$ to $s(x) \mathbin{\text{\S}} (F(T) * F(U))$. The merge operation $z$ transforms $(x, y) \mathbin{\$} (F(T) * F(U))$ to $z(x, y) \mathbin{\$} F(T * U)$. Connecting to category theory, a Functor that also satisfies SepHom is a lax-monoidal functor in the category constructed in §5.1, taking predicate separation conjunction as the tensor product.

SepHom is useful when partitions for different element components of a data container are referenced by different structures in a program. For example, in our case studies, $x \mathbin{\$} \text{Ref}_{addr}(T)$ is separating-homomorphic by SepHom($\text{Ref}_{addr}, \lambda x.\, x, \lambda x.\, x$). It allows us to split a reference to a composite memory object into references to each of its components. To illustrate, recall that we use $\{a{:}T\} * \{b{:}U\}$, abbreviated as $\{a{:}T, b{:}U\}$, to represent a record with two fields. The SepHom of Ref implies transformation $(x, y) \mathbin{\$} \text{Ref}_{addr}\{a{:}T, b{:}U\} \longrightarrow x \mathbin{\$} \text{Ref}_{addr}\{a{:}T\} * y \mathbin{\$} \text{Ref}_{addr}\{b{:}U\}$, which allows us to split one reference to the record into two references to each field of the record. Note, the SepHom of Ref does not hold on concrete memory models. Our system supports basic *fictional separation* to lift the concrete address-to-bytes memory model to an abstract one based on a map from records with fields to values with permissions, on which $addr \mapsto \{a{:}u, b{:}v\} = addr \mapsto \{a{:}u\} * addr \mapsto \{b{:}v\}$ holds and therefore the SepHom of Ref holds. These are detailed in §10.1.

Regarding generality, SepHom is also ubiquitous. $\text{Ref}_a$ is an example already introduced above. Permission modality and Later modality are also examples. We have SepHom($n\oplus, \lambda x.\, x, \lambda x.\, x$) and SepHom($\triangleright, \lambda x.\, x, \lambda x.\, x$). Another example is array, satisfying SepHom($\text{Array}_a$, unzip, zip), where $[x_1, \cdots, x_n] \mathbin{\$} \text{Array}_a(T) \triangleq (x_1 \mathbin{\$} \text{Ref}_a T) * \cdots * (x_n \mathbin{\$} \text{Ref}_{a+n-1} T)$ is defined as a collection of array elements connected by ($*$). List operations $\text{zip}([a_1, \cdots, a_n], [b_1, \cdots, b_n]) \triangleq [(a_1, b_1), \cdots, (a_n, b_n)]$, and $\text{unzip}[(a_1, b_1), \cdots, (a_n, b_n)] \triangleq ([a_1, \cdots, a_n], [b_1, \cdots, b_n])$ are defined as usual. The example in §6.6 illustrates how to utilize the SepHom of arrays in practice.

For more advanced data structures, SepHom might seem uncommon due to the presence of control structures that cannot be easily split and shared between divided abstractions. For instance, consider a dynamic array, which typically includes a length record. When attempting to separate dynamic-array($T * U$) into dynamic-array($T$) and dynamic-array($U$), a challenge arises: which separated part should own the length record? This issue suggests that a straightforward refinement like dynamic-array($T * U$) may not be separable. However, a solution exists through the *fiction of disjointness* [19, 35]. This technique allows us to 1) freeze the state of the common structure and 2) share constant copies of this frozen state between divided abstractions For example, in our dynamic array case, we could freeze the length and provide a copy of this constant length to each separated part of the array. Conclusively, it is possible to construct separating-homomorphic refinements for most data containers.

## 5.3 Modules over Rings

The next two algebraic properties are inspired by the theory of modules over rings. A left module-like structure over a ring-like structure $(\mathcal{S}, +, \cdot)$ comprises a group-like structure $(\mathcal{G}, *)$ and a scalar multiplication $\bullet : \mathcal{S} \times \mathcal{G} \to \mathcal{G}$. For any $n, m \in \mathcal{S}$ and $x, y \in \mathcal{G}$, a module-like structure may satisfy, (1) Scalar associativity, $(n \cdot m) \bullet x = n \bullet (m \bullet x)$; (2) Scalar distributivity, $(n + m) \bullet x = (n \bullet x) * (m \bullet x)$; (3) Distributivity over group operation, $n \bullet (x * y) = (n \bullet x) * (n \bullet y)$; (4) Scalar zero, $0 \bullet x = \epsilon$ for $0 \in \mathcal{S}$ and $\epsilon \in \mathcal{G}$; (5) Scalar identity, $1 \bullet x = x$ for $1 \in \mathcal{S}$.

Let us consider a predicate operator $F_n$ parameterized by a scalar $n$ belonging to some ring-like partial algebra $(\mathcal{S}, +, \cdot)$ called *scalar algebra*, ranged over by $n, m, \delta$. Construct a left module-like structure over $(\mathcal{S}, +, \cdot)$, such that the carrier $\mathcal{G}$ is the set of all SL predicates, the group operation ($*$) is predicate separating conjunction ($*$), the group identity is predicate $\text{Emp}(x) \triangleq \text{emp} \wedge x = ()$, and the scalar multiplication ($\bullet$) is the predicate operator $F$. Based on this module-like construction, properties **SA**, **SD**, **S1**, **S0** represent the laws of scalar associativity, scalar distributivity, identity, and scalar zero, respectively; the SepHom that specifies transformations between $F_n(T * U)$ and $F_n(T) * F_n(U)$, corresponds to the third axiom, Distributivity over the group operation.

Unlike Functor and SepHom, the module-like properties do not have a general interpretation in terms of data refinement. Instead, various refinements and modalities with distinct meanings can be specified by the language of modules. We present some examples in the following discussion.

(1) Permission modality $n \oplus T$ satisfies both Assoc($\oplus$, $\lambda n\,m\,x.\,x$, $\lambda n\,m\,x.\,x$), Dist($\oplus$, $\lambda n\,m\,x.\,(x,x)$, $\lambda n\,m\,(x,x).\,x$), SUnit($\oplus$, $\lambda \epsilon\,x.\,x$, $\lambda \epsilon\,x.\,x$), and SZero($\oplus$, $\top$), where $\lambda(x,x).\,x$ denotes a partial map from $(x,x)$ to $x$ for any $x$. The scalar algebra is a partial algebra obtained by restricting elements in the rational field to interval $(0, 1]$. However, if we extend the domain of permission to allow it to be locally greater than 1, e.g., to allow $\frac{1}{2} \oplus (2 \oplus T) = 1 \oplus T$, (the similar relaxation is also seen in [17]), then the scalar algebra extends to the semiring of non-negative rationals, and the module of $\oplus$ extends to a semimodule. Transformations of $n \oplus T$ then can be perfectly described by the laws of semimodule as follows:
- Scalar Distributivity, $x \mathbin{\texttt{\textsection}} (n+m) \oplus T \longleftrightarrow (x \mathbin{\texttt{\textsection}} n \oplus T) * (x \mathbin{\texttt{\textsection}} m \oplus T)$, for sharing of ownership.
- Scalar Associativity, $x \mathbin{\texttt{\textsection}} n \oplus m \oplus T \longleftrightarrow x \mathbin{\texttt{\textsection}} (n \cdot m) \oplus T$ states that $n$ proportion of $m$ proportion of ownership equals $n \cdot m$ proportion of ownership.
- The property of identity states unit permission can be omitted, $x \mathbin{\texttt{\textsection}} 1 \oplus T \longleftrightarrow x \mathbin{\texttt{\textsection}} T$
- The property of zero states zero permission means empty, $x \mathbin{\texttt{\textsection}} 0 \oplus T \longleftrightarrow$ emp.

(2) Let $x \mathbin{\texttt{\textsection}} \text{Path}_n T$ represent a subtree $x \mathbin{\texttt{\textsection}} T$ located at path $n$. Let the scalar algebra $(\mathscr{S}, \cdot)$ be the monoid of path concatenation and leave the scalar addition unspecified. Path satisfies Assoc(Path, $\lambda n\,m\,x.\,x$, $\lambda n\,m\,x.\,x$) and SUnit(Path, $\lambda n\,x.\,x$, $\lambda n\,x.\,x$). The properties characterize the path concatenation and empty path. This predicate operator Path is useful in cases where a tree abstraction is used, e.g., file systems. Particularly, in our case studies, the memory model of a nested record whose fields can be another nested record is represented by such a tree, whose edge labels are field names and leaves are non-aggregate values like integers. We use $x \mathbin{\texttt{\textsection}} \text{Path}_n T$ to represent a field $n$ that has a value refining $x$ w.r.t. $T$. The scalar $n$ represents a dot-connected path (e.g., a.b.c) that locates a member field in a nested record (e.g., {a: {b: {c: int}, d: int}, e: int}).

(3) Let $[l_i, \cdots, l_{i+k-1}] \mathbin{\texttt{\textsection}} \text{Slice}_{[i,i+k)} T$ specify the slice of an array $l$ that starts from index $i$ and has length $k$. Let the scalar algebra $(\mathscr{S}, +)$ be the partial semigroup of interval concatenation (defined as follows) and leave the multiplication unspecified.

$$[i,j) + [j,k) \triangleq [i,k), \quad \text{and} \quad [i,j) + [j',k) \text{ is undefined if } j \neq j'.$$

We also stipulate that any $[i, i)$ is a zero element and any $[i, i+1)$ is an identity element of the semigroup. Predicate operator Slice satisfies Dist(Slice, split, cat) and SZero(Slice, $\{[\,]\}$) where, assuming $l \mathbin{\texttt{\textsection}} \text{Slice}_{[i,i+k)} T$ maintains an invariant such that the length of $l$ equals $k$,

$\text{cat}_{[i,j),[j',k)}(l_1, l_2) \triangleq$ if $j = j'$ then the concatenation of $l_1, l_2$ else undefined.
$\text{split}_{[i,j),[j',k)}(l) \triangleq$ if $j = j'$ then $([l_0, \cdots, l_{j-i-1}], [l_{j-i}, \cdots, l_{k-i-1}])$ else undefined.

The scalar distributivity of array slices represents the concatenation and splitting of sub-slices. The property zero then specifies predicate application $l \mathbin{\texttt{\textsection}} \text{Slice}_{[i,i)} T$ equals empty because the interval is empty. Slice also has an extended identity property, SUnit'(Slice, Ref, $\lambda[x].\,x$, $\lambda x.\,[x]$) if we extend the definition **S1** to define,

$$\text{SUnit}'(F, G, f, g) \triangleq \text{for any } T, \text{ identity scalar } \epsilon \in \mathscr{S}, \text{ any } x \in \text{dom}(f_\epsilon) \text{ and } y \in \text{dom}(g_\epsilon), \quad \textbf{(S1')}$$
$$x \mathbin{\texttt{\textsection}} F_\epsilon(T) \longrightarrow f_\epsilon(x) \mathbin{\texttt{\textsection}} G_\epsilon(T) \text{ and } y \mathbin{\texttt{\textsection}} G_\epsilon(T) \longrightarrow g_\epsilon(y) \mathbin{\texttt{\textsection}} F_\epsilon(T) \text{ hold.}$$

The identity property states how to unwrap a slice into a single reference object when the slice specifies exactly one element.

(4) Finite separating quantifier $\{x_i\}_{i \in I} \mathbin{\texttt{\textsection}} \mathbin{\Large\text{*}}_{i \in I} T_i \triangleq (x_{i_1} \mathbin{\texttt{\textsection}} T_{i_1} * \cdots x_{i_n} \mathbin{\texttt{\textsection}} T_{i_n})$ for $I = \{i_1, \cdots, i_n\}$, is the fourth example of module-like predicate operators, where the scalar $I$ is a finite set. The addition between the scalars is the disjoint union $\sqcup$, and every singleton set is a scalar identity.

The zero is the empty set. $\circledast$ satisfies $\text{Dist}(\circledast, \text{disjoin}, \text{join})$ where $\text{disjoin}_{I,J}(\{x_k\}_{k \in I \sqcup J}) \triangleq (\{x_k\}_{k \in I}, \{x_k\}_{k \in J})$ and $\text{join}(x, y) \triangleq x \sqcup y$.

## 6 Utilizing the Algebraic Abstractions to Solve Transformation Problems

In this section, we introduce a generic SL reasoner designed to apply rules for transforming predicates from one form to another, directly mirroring the transformations described by the algebraic properties introduced in Section 5.

The reasoning rules are automatically instantiated from *templates* leveraging the generality provided by their parameters, which are algebraic properties. Collectively, they form a generic reasoning procedure capable of automating the reasoning for any predicate satisfying the specified properties. This endows our reasoner with a high degree of generalization capability.

### 6.1 Transformation Problem (TP) and bi-Abductive Transformation Problem (bi-TP)

Recall that formula $x \,\mathring{,}\, T \longrightarrow f(x) \,\mathring{,}\, U$ represents a transformation between refinements: any concrete construct refining $x$ w.r.t $T$ also refines $f(x)$ w.r.t $U$. Based on this interpretation, we define a *Transformation Problem (TP)* as:

*Definition 6.1 (TP).* Given predicates $T, U$ and a set $D$, a *Transformation Problem* $\text{TP}(T, U, D)$ looks for a pair $(\theta, f)$, where $\theta$ is an *FOL* formula that represents proof obligation, and $f$ is a function such that $\forall x \in D.\ x \,\mathring{,}\, T \longrightarrow f(x) \,\mathring{,}\, U$ holds if $(\theta \wedge D \subseteq \text{dom}(f))$ holds.

Intuitively, given any concrete construct $w$ refining abstraction $x$ w.r.t. refinement relation $T$, TP looks for the abstraction of $w$ under (another) refinement relation $U$. Formula $\theta$ is the proof obligation that entails the validity of the transformation. It constitutes a part of the final output of our SL reasoner — the proof obligation of program correctness.

However, a limitation is that $\text{TP}(T, U, D)$ only considers predicates $T, U$ that specify the same partitions of program states. In practice, we often encounter situations where predicates specify different partitions of states or different resource instances, and in that case, the partitions may only partially overlap. For example, consider three resources $A, B, C$. Let predicate $T$ specify $A, B$, while $U$ specify $B, C$. The transformation from $T$ to $U$ then cannot be described by TP, as it would be trivially unsolvable. This motivates us to introduce a variant of TP, inspired by [10].

*Definition 6.2 (bi-TP).* Given SL predicates $T, U$ and a set $D$ of terms, a *bi-abductive Transformation Problem* $\text{bi-TP}(T, U, D)$ looks for a quadruple $(\theta, f, Z, R)$, for an *FOL* formula $\theta$, a function $f$ and two predicates $Z, R$, such that $(\theta, f)$ is a solution of $\text{TP}(T*Z, U*R, D)$.

*Definition 6.3.* The operator $(*)$ between predicates is defined as $(T * Z)(x, z) \triangleq T(x) * Z(z)$.

Compared to TP, bi-TP additionally looks for two predicates, *frame R* and *anti-frame Z*, which represent the resources covered by $T$ but not by $U$ (in our example, $A$) and the resources covered by $U$ but not by $T$ (in our example, $C$). Intuitively, anti-frame $Z$ represents the missing resources not presented in $T$ but claimed by $U$; frame $R$ the remaining resources of $T$ after extracting $U$ from it. A solution $(\theta, f, Z, R)$ of $\text{bi-TP}(T, U)$ then states, in order to transform $T$ to $U$, some additional resources (specified by) $Z$ are required; the transformation also leaves $R$ as a remainder. Symbols $\theta$ and $f$ retain their meanings from TP.

For example, let $x \,\mathring{,}\, \{a\colon T\}$ specify a record with one field $a$ whose value refines $x$ w.r.t. $T$. Also let $\{a\colon T_1\} * \{b\colon T_2\}$, abbreviated as $\{a\colon T_1, b\colon T_2\}$, specify a record with two fields $a, b$. $\text{bi-TP}(\{a\colon T_1, b\colon T_2\}, \{b\colon T_2', c\colon T_3\}, D)$ for some $D$ is an example for partial overlapping in field b between source and target predicates. The bi-TP has a solution $(\theta, \lambda((a, b), c).\,((f(b), c), a), \{c\colon T_3\}, \{a\colon T_1\})$ for some $\theta, f$. This tells us how to transform abstractions from $T_2$ to $T_2'$ (using $f$), and indicates that $\{c\colon T_3\}$ is needed to complete the transformation, and $\{a\colon T_1\}$ stays available for future transformations.

## 6.2 A Generic TP/bi-TP Solver

Let $\mathfrak{R}$ be a sequence of inference rules. The solver is a typical backward reasoning procedure.

*Notation* ($\mathscr{P} \leftarrow \mathscr{A}$). For a problem $\mathscr{P}$ that can be a TP, a bi-TP, or a bi-EP (introduced later), $\mathscr{P} \leftarrow \mathscr{A}$ denotes a judgment, "problem $\mathscr{P}$ has a solution $\mathscr{A}$".

**Routine 1** (TP/bi-TP Solver). Given a problem $\mathscr{P}$, which can be a TP$(T, U, D)$ or a bi-TP$(T, U, D)$, Routine 1 performs backward reasoning using rules in $\mathfrak{R}$ to generate a proof tree with a root $\mathscr{P} \leftarrow \mathscr{A}$ for some $\mathscr{A}$. The rules are prioritized; one that occurs earlier in $\mathfrak{R}$ has a higher priority. This means that, whenever more than one rule is applicable to a subgoal, the routine adopts the rule that occurs earliest in $\mathfrak{R}$. It ensures the generated proof tree is unique. Let $\mathscr{P} \leftarrow \mathscr{A}$ denote the root of the unique tree. If all its leaves are axioms, return $\mathscr{A}$. Otherwise, the routine fails.

Specifically, $\mathfrak{R}$ consists of three sorts of rules: ad-hoc rules, rules instantiated from templates, and fallbacks. The rules are prioritized: ad-hoc rules have the highest priority, then instantiated rules, and at last fallbacks. This priority scheme allows users to override instantiated rules by providing custom ad-hoc rules. This enables fine-tuning in specific cases where the generic reasoning procedure derived from the templates may not perform optimally. In the following subsections, we introduce the three sorts of rules, respectively, and focus most of our attention on the templates.

## 6.3 Ad-hoc Rules

Ad-hoc rules are used to cover the special cases that cannot be handled by the generic templates over general algebraic abstractions. In our experience, the special cases are limited to:

(1) Specific transformation problems that cannot be generally specified by algebraic axioms, e.g., the one reinterpreting arrays of structures to structures of arrays. As another example, TPs and bi-TPs between identical refinement relations are solved by the following ad-hoc rules.

$$\frac{\text{Axiom}}{\text{TP}(T, T) \leftarrow (\text{true}, \lambda x.\, x)} \qquad \frac{\text{Axiom}}{\text{bi-TP}(T, T) \leftarrow (\text{true}, \lambda x.\, x, \text{Emp}, \text{Emp})} \text{ (ID)}$$

Emp$(x) \triangleq \text{emp} \wedge x = ()$ is the refinement relation for empty.

(2) Elimination of predicate counterparts of logic connectives. For example, our case studies require two rules to eliminate predicate $(*)$ defined as $(T_1 * T_2)(x_1, x_2) \triangleq T_1(x_1) * T_2(x_2)$. One rule reduces bi-TP$((T_1 * T_2), U, D)$ to bi-TP$(T_1, U, D_1)$ and bi-TP$(T_2, Z', D_2)$ for certain $Z', D_1, D_2$. Another rule reduces bi-TP$(T, (U_1 * U_2), D)$ to bi-TP$(T, U_1, D_1)$ and bi-TP$(R', U_2, D_2)$ for certain $R', D_1, D_2$. We leave the details to Appendix C, in order to prevent distracting readers.

## 6.4 Fallbacks

The rules that have the lowest priorities are fallbacks. When a bi-TP$(T, U, D)$ fails to be covered by either ad-hoc rules or instantiated rules, fallbacks are applied, which are listed as follows.

$$\frac{\text{TP}(T, U, D) \leftarrow (\theta, f)}{\text{bi-TP}(T, U, D) \leftarrow (\theta, f, \text{Emp}, \text{Emp})} \text{ (FB}_1) \qquad \frac{\text{Axiom}}{\text{bi-TP}(T, U, D) \leftarrow (\text{true}, (\lambda(x, z).\, (z, x)), U, T)} \text{ (FB}_2)$$

The first fallback reduction (FB$_1$) downgrades a bi-TP to a TP to allow the reasoning process to attempt rules and templates written for the TP form (e.g., the transformation between rationals and integer pairs and transformations of predicates that are not separating-homomorphic). If the TP$(T, U, D)$ induced by the first fallback fails to be solved, given that there is no further fallback for TP, the algorithm attempts the second fallback (FB$_2$). The fallback considers the partitions of program states specified by $T$ and $U$ to be disjoint. Therefore, the fallback assigns the demand for transforming $T$ to $U$ to be the entire $U$ and the remainder to be the entire $T$.

$$\text{given SZero}(F, D'), \quad \frac{\text{bi-TP}(\text{Emp}, U, \{()\}) \leftarrow (\theta, f, Z, R)}{\text{bi-TP}(F_0(T), U, D) \leftarrow (\theta \wedge D \subseteq D', f, Z, R)} \text{ (S0}_\text{L}\text{)}$$

$\{()\}$ is the singleton set containing only $()$. $x \mathbin{\raisebox{0.2ex}{\scalebox{0.6}{$\circ$}}} \text{Emp} \triangleq \text{emp} \wedge x = ()$.

$$\text{given SZero}(F, D'), \quad \frac{\text{bi-TP}(T, \text{Emp}, D) \leftarrow (\theta, f, Z, R)}{\text{bi-TP}(T, F_0(U), D) \leftarrow (\theta \wedge y \in D', \mathfrak{m}_1(\lambda\_.y) \circ f, \text{Emp}, U)} \text{ (S0}_\text{R}\text{)}$$

$$\text{given Functor}(F, m, d), \quad \frac{\text{TP}(T, U, D \gg\!\!= d) \leftarrow (\theta, f)}{\text{TP}(F(T), F(U), D) \leftarrow (\theta, m(f))} \text{ (TF)}$$

recalling $(D \gg\!\!= d) \triangleq \bigcup_{x \in D} d(x)$ is the monadic bind of sets

$$\text{given Functor}(F, m, d) \text{ and SepHom}(F, s, z), \quad \frac{\text{bi-TP}(T, U, D \gg\!\!= (d \circ z)) \leftarrow (\theta, f, Z, R)}{\text{bi-TP}(F(T), F(U), D) \leftarrow (\theta, s \circ m(f) \circ z, F(Z), F(R))} \text{ (SH)}$$

$$\text{given Dist}(F, s, z), \quad \frac{\text{bi-TP}(F_m(T), F_m(U), h(D)) \leftarrow (\theta, f, Z, R)}{\text{bi-TP}(F_n(T), F_m(U), D) \leftarrow (\theta, f \circ h, F_\delta(T) * Z, R)} \text{ (SD}_\text{L}\text{)} \quad \text{if } n \neq m \text{ and } n + \delta = m$$
where $h = \lambda(x_n, (x_\delta, w)). (z_{n,\delta}(x_n, x_\delta), w);$

$$\text{given Dist}(F, s, z), \quad \frac{\text{bi-TP}(F_m(T), F_m(U), h(D)) \leftarrow (\theta, f, Z, R)}{\text{bi-TP}(F_n(T), F_m(U), D) \leftarrow (\theta, g, Z, F_\delta(T) * R)} \text{ (SD}_\text{R}\text{)} \quad \text{if } n \neq m \text{ and } n = m + \delta$$
where $h = \lambda(x_n, w). \text{let } (x_m, x_\delta) = s_{m,\delta}(x_n) \text{ in } (x_m, w)$
$g = \lambda(x_n, w). \text{let } (x_m, x_\delta) = s_{m,\delta}(x_n); (y, r) = f(x_m, w) \text{ in } (y, (x_\delta, r))$

$$\text{given Assoc}(F, g, h), \quad \frac{\text{bi-TP}(F_m(F_\delta(T)), F_m(U), \mathfrak{m}_1(h_{m,\delta})(D)) \leftarrow (\theta, f, Z, R)}{\text{bi-TP}(F_n(T), F_m(U), D) \leftarrow (\theta, f \circ \mathfrak{m}_1(h_{m,\delta}), Z, R)} \text{ (SA}_\text{L}\text{)} \quad \text{if } n \neq m \text{ and } n = m \cdot \delta$$

$$\text{given Assoc}(F, g, h), \quad \frac{\text{bi-TP}(F_n(T), F_n(F_\delta(U)), D) \leftarrow (\theta, f, Z, R)}{\text{bi-TP}(F_n(T), F_m(U), D) \leftarrow (\theta, \mathfrak{m}_1(g_{n,\delta}) \circ f, Z, R)} \text{ (SA}_\text{R}\text{)} \quad \text{if } n \neq m \text{ and } n \cdot \delta = m$$

$$\text{given SUnit}(F, g, h), \quad \frac{\text{bi-TP}(F_\epsilon(T), F_m(U), D) \leftarrow (\theta, f, W, R)}{\text{bi-TP}(T, F_m(U), D) \leftarrow (\theta, \mathfrak{m}_1(h_\epsilon) \circ f, W, R)} \text{ (S1}_\text{I}\text{)} \quad \text{if } T \text{ does not match pattern } F_n(T') \text{ for any } n, T'$$

$$\text{given SUnit}(F, g, h), \quad \frac{\text{bi-TP}(T, U, \mathfrak{m}_1(g_\epsilon)(D)) \leftarrow (\theta, f, W, R)}{\text{bi-TP}(F_\epsilon(T), U, D) \leftarrow (\theta, f \circ \mathfrak{m}_1(g_\epsilon), W, R)} \text{ (S1}_\text{E}\text{)} \quad \text{if } U \text{ does not match pattern } F_m(U') \text{ for any } m, U'$$

**Notation** $(\mathscr{P} \leftarrow \mathscr{A})$ *denotes a judgment, "problem $\mathscr{P}$ has a solution $\mathscr{A}$".*
Notation $h(D)$ *denotes the image of set $D$ under function $h$. Mapper $\mathfrak{m}_1(h) \triangleq \lambda(x, y). (h(x), y)$.*

Fig. 3. Representative templates of rules for TPs and bi-TPs. Parameters of the templates are notated by the "given" clauses on the left side. Side conditions are given on the right side.

### 6.5 Generic Templates Parameterized by Algebraic Properties

The reasoner uses rules instantiated from templates shown in Fig. 3 to solve TPs and bi-TPs. A template takes a set of algebraic properties as its parameters. Given a set of instances of these properties, the template is instantiated into a concrete reasoning rule by (1) substituting the instance arguments for the property parameters throughout the template; (2) deriving the proof of the instantiated rule using the proofs of the instances. Instantiated reasoning rules are then inserted into $\mathfrak{R}$ so the reasoner can use it. Each template defines a reduction aimed at eliminating a predicate operator, and they are organized according to the priority of the rules they instantiate.

Rules instantiated from S0$_\text{L}$ and S0$_\text{R}$ are attempted first, eliminating any zero-parameterized operator $F_0(T)$ for the left side of the bi-TP, and $F_0(U)$ for the right side, respectively. They reduce the given bi-TP to a form solvable by the ad-hoc rules of Emp (cf. Appendix C).

Next, the reasoner tries to apply TF and SH. SH is essentially the bi-abductive version of TF. They both depend on property TF, and SH additionally demands property SH. Given a TP($F(T),F(U),D$) or a bi-TP($F(T), F(U), D$), they eliminate the common predicate operator $F$. Intuitively, if $F(T)$ represents a container, the templates forward the reasoning process from the container's space to its elements' space, reducing the problem about the container to a problem about its elements.

If we consider $F_n(T), F_m(U)$ as slices of some structure, templates $SD_L$ and $SD_R$ leverage the scalar distributivity property **SD** to split and merge slices. Given bi-TP($F_n(T), F_m(U), D$), by checking if $n = m + \delta$, $SD_R$ checks if the source slice $F_n(T)$ covers a larger domain than the target $F_m(T)$, represented by $F_\delta(T)$. If so, it splits the source slice into two: $F_m(T)$ and $F_\delta(U)$. Then it leaves the slice $F_\delta(T)$ as a residue and induces a subproblem bi-TP($F_m(T), F_m(U), D$) reducible by SH.

The case in $SD_L$ is symmetric to $SD_R$, but checks if the source slice cannot cover the target and demands an additional $\delta$-portion. Let us explain this trying to solve bi-TP(Slice$_{[i,j)}$ $T$, Slice$_{[i,k)}$ $U, D$). If $j < k$ it implies $[i, j) + [j, k) = [i, k)$. Thus, $SD_L$ is applicable if we instantiate $n, \delta, m$ to $[i, j), [j, k), [i, k)$. Application of $SD_L$ induces the subgoal bi-TP(Slice$_{[i,k)}$ $T$, Slice$_{[i,k)}$ $U, h(D)$) and indicates that Slice$_{[j,k)}(T)$ is demanded to be extracted in the subsequent reasoning process, where $h(D)$ augments the Domain of $T$ with $\delta$.

Templates $SD_L$ and $SD_R$ are insufficient when the scalar addition is not commutative, associative, and cancellation; for example, interval addition that is non-commutative. If $j < k$ and $i' < i$, bi-TP(Slice$_{[i,j)}$, Slice$_{[i',k)}, D$) cannot be handled by $SD_L$ and $SD_R$ because there is no $\delta$ such that $[i, j) + \delta = [i', k)$ or $[i, j) = [i', k) + \delta$. Instead, it requires a template for ($\delta' + n + \delta = m$) and instantiates $\delta', n, \delta, m$ to $[i', i), [i, j), [j, k), [i', k)$. This is detailed in Appendix F.

Templates $SA_L$ and $SA_R$ leverage property **SA** to respectively collapse nested $F_n(F_m(T))$ into $F_{n \cdot m}(T)$ and expand collapsed $F_{n \cdot m}(T)$ into $F_n(F_m(T))$. They are symmetric to $SD_L$ and $SD_R$ in terms of considering scalar multiplication instead of scalar addition,

Notably, our reasoner does *not* generically support module-like operators that are both scalar associative and scalar distributive. Consider a bi-TP($\frac{1}{2} \circledcirc T, \frac{1}{3} \circledcirc U, D$) and assume $T = \frac{1}{4} \circledcirc T'$ for some $T'$. Permission modality $\circledcirc$ is both scalar associative and scalar distributive. There are indeterminate solutions $\delta, \delta'$ such that $\frac{1}{2} = \delta \cdot \frac{1}{3} + \delta'$. Consequently, we do not know which value should be used to instantiate $\delta$ in $SA_L$, $SA_R$, $SD_L$, or $SD_R$. However, it is possible to rewrite $F_n(F_m(T))$ to $F_{n \cdot m}(T)$ for any scalar-associative $F$ and any $n, m, T$ before invoking the TP/bi-TP solver. If after this it is not necessary to apply $SA_L$ and $SA_R$, the reasoning succeeds by applying $SD_L$ and $SD_R$.

Finally, if a bi-TP($T, F_b(U), D$) expects to transform a non-module-like $T$ to a module-like $F_b(U)$, $S1_I$ wraps $T$ to $F_\epsilon(T)$. Conversely, if a bi-TP($F_\epsilon(T), U, D$) expects to transform a module-like $F_\epsilon(T)$ to a non-module-like $U$, $S1_E$ unwraps $F_\epsilon(T)$ when its scalar $\epsilon$ is an identity.

As a side note, the side conditions in the templates are arithmetic equations within the ring-like scalar algebra(s). We assume our reasoner is parameterized by solvers to handle them.

### 6.6 Example: Matrix Partitioning

To illustrate how instantiated rules from the templates are applied to real problems, we consider the transformation that splits the abstraction of a big matrix $\begin{pmatrix} A & B \\ C & D \end{pmatrix}$ into four sub-matrixes $A, B, C, D$, which is a key step in our case study of Strassen's matrix multiplication algorithm.

A $2N \times 2N$ matrix is represented by a two-dimensional array of lengths $2N$. We represent its refinement by predicate Slice$_{[0,2N)}$(Slice$_{[0,2N)}$ $\mathbb{Z}$). The desired transformation is then

$$l \, \S \, \text{Slice}_{[0,2N)}(\text{Slice}_{[0,2N)} \, \mathbb{Z}) \longrightarrow \begin{pmatrix} l_A \, \S \, \text{Slice}_{[0,N)}(\text{Slice}_{[0,N)} \, \mathbb{Z}) & * & l_B \, \S \, \text{Slice}_{[0,N)}(\text{Slice}_{[N,2N)} \mathbb{Z}) & * \\ l_C \, \S \, \text{Slice}_{[N,2N)}(\text{Slice}_{[0,N)} \, \mathbb{Z}) & * & l_D \, \S \, \text{Slice}_{[N,2N)}(\text{Slice}_{[N,2N)} \, \mathbb{Z}) \end{pmatrix}$$

where $l_A = l_{[0:N][0:N]}$, $l_B = l_{[0:N][N:2N]}$, $l_C = l_{[N:2N][0:N]}$ and $l_D = l_{[N:2N][N:2N]}$, in Python's slicing notation. The transformation raises four bi-TPs for extracting each sub-matrix. Let us consider the first bi-TP, ignore the domain of the bi-TP, and leave the detailed reduction process to Appendix D,

$$\text{bi-TP}(\, \text{Slice}_{[0,2N)}(\text{Slice}_{[0,2N)} \, \mathbb{Z}) \,, \, \text{Slice}_{[0,N)}(\text{Slice}_{[0,N)} \, \mathbb{Z}), \cdots). \tag{1}$$

Generically Automating Separation Logic by Functors, Homomorphisms, and Modules 67:15As Slice satisfies **SD**, this property of Slice instantiates template $\text{SD}_\text{R}$. The instantiated rule is shown in Appendix D. The rule reduces goal (1) to the following bi-TP.

$$\text{bi-TP}(\ \text{Slice}_{[0,N)}(\text{Slice}_{[0,2N)}\ \mathbb{Z})\ ,\ \text{Slice}_{[0,N)}(\text{Slice}_{[0,N)}\ \mathbb{Z})\ ,\ \cdots\ ) \quad (2)$$

As Slice also satisfies **TF** and **SH**, these properties of Slice instantiate template **SH**. The instantiated rule eliminates the outer-most common operator $\text{Slice}_{[0,N)}$ and reduces goal (2) to,

$$\text{bi-TP}(\ \text{Slice}_{[0,2N)}\ \mathbb{Z}\ ,\ \text{Slice}_{[0,N)}\ \mathbb{Z}\ ,\ \cdots\ ) \quad (3)$$

Again, our reasoner applies the previous rule instantiated from $\text{SD}_\text{R}$ to split the source slice and to reduce goal (3) to bi-TP( $\text{Slice}_{[0,N)}\ \mathbb{Z}$, $\text{Slice}_{[0,N)}\ \mathbb{Z}$, $\cdots$) which is immediately solvable by (ID) rule.

## 7 Programming Language and wp-Transformer

Starting from this section, we turn to complete our SL reasoner. We demonstrate how a program verifier can be built on top of the TP/bi-TP solver, in order to show that algebra-based rule generation can benefit program verification. To ground our discussion on program verification, and in line with our automation algorithm's generic design, we base our discussion on a minimally-specified generic formalization that can be instantiated to many concrete languages, including C.

### 7.1 A Generic Formalization for Programming Languages

We formalize our language using non-deterministic state monad to define a generic semantic formalization for programming languages. We abstract the language operations as a set of operators *Opr* represented by $\rho$. Operators can be parameterized by one or more programs, becoming a higher-order operator, with $\text{arity}(\rho)$ denoting the number of higher-order parameters of $\rho$. This allows for the formalization of control flow (e.g., If and While) as higher-order operators.

$u, v \in \textit{Value},\quad \textit{Program} \subseteq \textit{Value} \to \text{Monad},\quad \text{where Monad} \triangleq \textit{State} \to \text{powerset}(\textit{Value} \times \textit{State})$

$\textit{Program} \ni C_1, C_2, \cdots ::= \rho(C_1, \cdots, C_{\text{arity}(\rho)}) \mid \lambda v.\ (C_1(v) \ggg C_2),\quad \text{where } (\ggg) \text{ is monadic bind}$

To illustrate how higher-order operators represent control flows, $\text{If}(C_\text{T}, C_\text{F})(b) \triangleq \text{if } b \text{ then } C_\text{T} \text{ else } C_\text{F}$ is an operator having $b$ as its argument and $C_\text{T}, C_\text{F}$ as two programs that represent the two branches.

### 7.2 Separation Logic over the Programming Language Formalization

Conventionally, we use $\mathbf{wp}_{C(u)}\{v.\ \psi(v)\}$ to denote the weakest precondition of a computation $C(u)$, where $\psi(v)$ is an SL formula parameterized by variable $v$. Judgement $(\mathbf{wp}_{C(u)}\{v.\ \psi(v)\} \dashv \phi)$ specifies that given argument $u$ and an initial state $s$ satisfying $\phi$, the computation $C(u)$ returns a value $v$ and results in a state $s'$ satisfying $\psi(v)$ — recall the discussion in §4 about our "big PCM".

$$(\mathbf{wp}_{C(u)}\{v.\ \psi(v)\} \dashv \phi) \triangleq \forall s\, s'\, v.\ (s \models \phi) \land (v, s') \in C(u) \longrightarrow (s' \models \psi(v)) \quad (\mathbf{wp})$$

Hoare triples are conventionally specified based on it, $\{\phi\}C(u)\{v.\ \psi(v)\} \triangleq (\mathbf{wp}_{C(u)}\{v.\ \psi(v)\} \dashv \phi)$.

Based on this generic formalization for programming languages, we do not stipulate any specific **wp** rule. Instead, we assume a set $\mathcal{W}$ of **wp** rules in the following form

$$\frac{\mathbf{wp}_{C_1(u_1)}\{v_1.\ \psi_1(v_1)\} \dashv \phi_1 \quad \cdots \quad \mathbf{wp}_{C_n(u_n)}\{v_n.\ \psi_n(v_n)\} \dashv \phi_n}{\mathbf{wp}_{\rho(C_1,\cdots,C_n)(u_0)}\{v_0.\ \psi_0(v_0)\} \dashv \phi_0} \text{ (\textbf{wp}-rule)} \quad \text{for fresh fixed variables } u_1, \cdots, u_n$$

Some examples are listed in Fig. 4. Based on the set $\mathcal{W}$, we formalize a standard **wp**-transformer.

**Routine 2 (wp-transformer).** Given $(\phi, C(u), \psi(v))$, perform backward reasoning using rules in $\mathcal{W}$ and rule (Bind) in Fig. 4 to generate a proof tree with a root $\mathbf{wp}_{C(u)}\{v.\ \psi(v)\} \dashv \phi'$ for some $\phi'$. Return $\phi \longrightarrow \phi'$ if all leaves in the tree are axioms. Otherwise, the routine fails.

Proc. ACM Program. Lang., Vol. 9, No. POPL, Article 67. Publication date: January 2025.

$$\frac{\mathbf{wp}_{C_2(u')}\{v.\,\psi(v)\} \dashv \psi'(u') \qquad \mathbf{wp}_{C_1(u)}\{v.\,\psi'(v)\} \dashv \psi}{\mathbf{wp}_{(C_1 \ggg C_2)(u)}\{v.\,\psi(v)\} \dashv \psi} \text{ (Bind)} \text{ for a fresh fixed variable } u'$$

$$\mathbf{wp}_{\text{load}(addr)}\{v.\,\psi(v)\} \dashv (x \, \S \, \text{Ref}_{addr} T) * \forall v.\, (x \, \S \, \text{Ref}_{addr} T * x \, \S \, \text{val}_v T \twoheadrightarrow \psi(v))$$

$$\mathbf{wp}_{\text{store}(addr,u)}\{v.\,\psi(v)\} \dashv (x \, \S \, \text{Ref}_{addr} T) * (y \, \S \, \text{val}_u U) * (y \, \S \, \text{Ref}_{addr} U \twoheadrightarrow \psi())$$

$$\frac{\mathbf{wp}_{C_T()}\{v.\,\psi(v)\} \dashv \phi_1 \qquad \mathbf{wp}_{C_F()}\{v.\,\psi(v)\} \dashv \phi_2}{\mathbf{wp}_{\text{If}(C_T,C_F)(u)}\{v.\,\psi(v)\} \dashv (P \, \S \, \text{val}_u \text{Bool}) * ((P \to \phi_1) \land (\neg P \to \phi_2))}$$

**Notations:** $(x \, \S \, \text{Ref}_{addr} T) \triangleq (\exists v.\, addr \mapsto v \land v \Vdash x \, \S \, T)$ claims the ownership of a memory object at address $addr$ and asserts it has a value $v$ refining $x$ w.r.t. $T$.
$(x \, \S \, \text{val}_v T) \triangleq (\text{emp} \land v \Vdash x \, \S \, T)$ asserts a value $v$ that refines $x$ w.r.t. $T$.

Fig. 4. Example rules for **wp**-transformer, using the simple imperative language IMP.

## 8 Connecting the TP/bi-TP Solver to Program Verification

In this section, we build an SL reasoner on top of the TP/bi-TP solver introduced in §6 to show that our algebraic method for TPs/bi-TPs can be used to tackle real-world program verification problems, which then implies the significance of our algebraic abstractions (§5) and automatic rule generation (§6) based on the algebraic abstractions.

This SL reasoner is based on a standard process using **wp**-transformer. Illustrated in Fig. 1, the process first uses the **wp**-transformer shown in Routine 2 to extract an SL entailment that implies the desired program correctness; then, it applies an SL entailment reasoner to extract an *FOL* formula as a proof obligation that implies the validity of the given SL entailment. The proof obligation is either sent to users for manual proof or to ATPs for automatic proof.

Focusing on proving SL entailments, Section 8.1 first restricts the domain of formulas considered by our reasoner. Then, as an intermediate step, §8.2.1 first reduces the decision problem of a given SL entailment to so-called *bi-abductive Entailment Problems* (bi-EPs), used merely as a stepping-stone. §8.2.2 finally reduces these bi-EPs to a series of bi-TPs. Through these steps, we establish a connection between our TP/bi-TP solver (§6) and program verification, completing the chain from program specifications to algebraic problem-solving. In the end of the section, §8.3 clarifies how the reasoning process matches hypotheses of an entailment with its goals; §8.4 provides details about how to instantiate existential quantifications and evars.

### 8.1 Restricting Formulas Reducible to bi-TPs

Before delving into formalizing the SL reasoner, we must clarify that not all decision problems of SL entailments are reducible to bi-TPs by our reasoner. Following RefinedC [64] and Argon [65], we restrict the formulas for SL entailments and program state specification to subsets **E** and **S**,

| | State | $\mathbf{S} ::= x \, \S \, T \mid \top \mid \bot \mid \text{emp} \mid \mathbf{S} * \mathbf{S} \mid \mathbf{S} \land P \mid \exists \alpha.\, \mathbf{S}$ | (S) |
|---|---|---|---|
| | Goal | $\mathbf{G} ::= \mathbf{S} \mid \mathbf{S} * \mathbf{G} \mid \mathbf{S} \twoheadrightarrow \mathbf{G} \mid P \to \mathbf{G} \mid \mathbf{G} \land \mathbf{G} \mid \forall \alpha.\, \mathbf{G} \mid \exists \alpha.\, \mathbf{G}$ | (G) |
| | Entailment | $\mathbf{E} ::= \mathbf{S} \to \mathbf{G}$ | (E) |

where $P$ ranges over *FOL* formulas; $T$ ranges over all SL predicates, *no matter if the formula of the definition of $T$ is within **E**, **G**, or **S**.* Specifically, this restriction means that, (1) we only consider program specifications $\{\phi\}C\{v.\,\psi(v)\}$ such that $\phi \in \mathbf{S}$ and $\psi(v) \in \mathbf{G}$; (2) for any **wp** rule given in the set $\mathscr{W}$ (cf., §7.2) and in form (wp-rule), we require $\psi_i(v_i) \in \mathbf{G}$ for every $i \in \{0, 1, \cdots, n\}$.

LEMMA 8.1 (**wp**-TRANSFORMER ROUTINE 2 IS CLOSED IN **E**). *For any program $C$, any $\phi \in \mathbf{S}$ and $\psi(v) \in \mathbf{G}$, the return of the **wp**-transformer Routine 2 belongs to **E**.*

Although the restriction on formulas limits the capability of our reasoner, a practical mitigation exists. To specify program states using a formula $\phi \notin \mathbf{S}$, users can define a predicate $T(x) \triangleq \phi$ to wrap $\phi$. The resulting formula $(x \mathbin{\S} T)$ belongs to $\mathbf{S}$. As long as users can provide algebraic properties of $T$, $\phi$ can be equivalently handled by our reasoner through rewriting $\phi$ into $x \mathbin{\S} T$. For example, consider the predicate definition $x \mathbin{\S} \mathrm{Ref}_a(T) \triangleq \exists v. (a \mapsto v) \wedge (v \Vdash x \mathbin{\S} T)$, which involves the satisfaction operator $\Vdash$ that is not in $\mathbf{S}$. Our reasoner can still handle $x \mathbin{\S} \mathrm{Ref}_a(T)$ because the algebraic properties of $\mathrm{Ref}_a(T)$ are provided. Through this wrapping technique, our reasoner can support formulas that would otherwise exceed $\mathbf{S}$ or $\mathbf{G}$. Consequently, our reasoner still targets the entire SL, provided that the necessary properties of the wrapper predicates are supplied.

## 8.2 Reduction from SL Entailments to bi-TPs

Given an SL entailment between two formulas, if any of the formulas is composite, we apply a decomposition process to reduce the decision problem of the entailment to decision problems of a series of smaller entailments between atomic formulas. We say a formula is *atomic* iff it is a *predicate application*, and *composite* iff it is not atomic ($\top, \bot, \mathrm{emp}$ are nullary connectives which are eliminated by the decomposition process). We want this decomposition because entailments between predicate applications, $x \mathbin{\S} T \longrightarrow y \mathbin{\S} U$, are close to the statement of bi-TP, allowing us to easily reduce their decision problems to bi-TPs.

Roughly, this decomposition splits a big entailment between $n$ source items $\phi_1 * \cdots * \phi_n$ and $m$ target items $\psi_1 * \cdots * \psi_m$ into about $n \times m$ smaller entailments between (parts of) $\phi_i$ and (parts of) $\psi_j$ for $1 \leq i \leq n$ and $1 \leq j \leq m$. Let us consider an entailment between $\phi$ and $\psi$ as extracting a bunch of target resources $\psi$ from a bunch of source resources $\phi$. The big entailment $\phi_1 * \cdots * \phi_n \longrightarrow \psi_1 * \cdots \psi_m$ then aims to extract the $m$ bunches of resources from the $n$ bunches. The decomposition divides the extraction into many small steps: We first extract the bunch $\psi_1$ from the bunch $\phi_1$. Because the two bunches may not be perfectly matched but partially overlapped, some part $Z_1$ of the target $\psi_1$ may be missing in the source $\phi_1$, and some part $R_1$ of the source $\phi_1$ may remain after the extraction. To find the missing $Z_1$, we move to the next bunch $\phi_2$, which may provide some components but still lack others, so we continue to look in $\phi_3, \cdots, \phi_n$. All the source bunches leave some parts $R_1, \cdots, R_n$ (can be empty) after the extraction. After $\psi_1$ is gathered, we turn to extract the next target $\psi_2$ from the remaining parts $R_1 * \cdots * R_n$ of the source bunches, in the same way described above. Repeatedly, we extract $\psi_3, \cdots, \psi_m$ iteratively from the remaining parts of the source bunches after each extraction.

To formalize the extraction that can leave some remaining source (e.g., $R_1$ above) and unfulfilled target (e.g., $Z_1$ above), we introduce *bi-abductive entailment*, $\phi * Z \longrightarrow \psi * R$, where *frame* $R$ represents the remaining part of $\phi$ after the extraction and *anti-frame* $Z$ the missing part claimed by $\psi$ but not seen in $\phi$. Both $Z, R$ are variables that represent unknowns to infer. We define the problems of inferring such $Z, R$ as *bi-abductive Entailment Problems*

*Definition 8.2 (bi-EP).* Given $\phi, \psi \in \mathbf{S}$, if $\exists$ does not occur in $\phi$, *a bi-abductive Entailment Problem* bi-EP$(\phi, \psi)$ looks for a triple $(\theta, Z, R)$, for an *FOL* formula $\theta$ as the proof obligation, and two SL formulas $Z, R \in \mathbf{S}$, such that $\phi * Z \longrightarrow \psi * R$ holds if $\theta$ holds.

The decomposition of composite entailments is realized by rules in Fig. 7. Rule ($*_\mathrm{L}$) and ($*_\mathrm{R}$) formalize the intuition at the beginning of this subsection: to extract $\psi$ from two bunches $\phi_1 * \phi_2$ of resources, we first look in bunch $\phi_1$. It may leave some unfulfilled target $Z$, so we continue to look for it in bunch $\phi_2$. The remainders $R_1, R_2$ of the two small extractions are left as the remainder of the original big extraction. In ($*_\mathrm{R}$), where we extract two bunches $\psi_1, \psi_2$ from one bunch $\phi$, we first extract $\psi_1$ from $\phi$ and then $\psi_2$ from the remaining part $R$ of $\phi$.

$$\frac{\text{bi-EP}(S_1, S_2) \leftarrow (\theta, S_Z, S_R) \quad \text{emp} \longrightarrow S_Z \quad S_R \longrightarrow \text{emp}}{\theta \mid S_1 \vdash S_2} \text{(bi-EP)} \quad \exists \text{ does not occur in } S_1$$

$$\frac{\text{bi-EP}(S_1, S_2) \leftarrow (\theta_1, S_Z, S_R) \quad \text{emp} \longrightarrow S_Z \quad \theta_2 \mid S_R \vdash G}{\theta_1 \land \theta_2 \mid S_1 \vdash S_2 * G} \text{(bi-EP}_R) \quad \exists \text{ does not occur in } S_1$$

$$\frac{\theta(\beta) \mid S(\beta) \vdash G}{\forall \alpha. \theta(\alpha) \mid \exists \alpha. S(\alpha) \vdash G} \text{(S}_\exists) \quad \text{for a fresh fixed variable } \beta \qquad \frac{\theta(\beta) \mid S \vdash G(\beta)}{\forall \alpha. \theta(\alpha) \mid S \vdash \forall \alpha. G(\alpha)} \quad \text{for a fresh fixed variable } \beta$$

$$\frac{\theta \mid S_1 * S_2 \vdash G}{\theta \mid S_1 \vdash S_2 \twoheadrightarrow G} \qquad \frac{\theta \mid S \land P \vdash G}{\theta \mid S \vdash P \rightarrow G} \qquad \frac{\theta_1 \mid S \vdash G_1 \quad \theta_2 \mid S \vdash G_2}{\theta_1 \land \theta_2 \mid S \vdash G_1 \land G_2} \qquad \frac{\theta(\beta) \mid S \vdash G(\beta)}{\forall \alpha. \theta(\alpha) \mid S \vdash \exists \alpha. G(\alpha)} \quad \text{for a fresh free variable } \beta$$

Fig. 5. Rules Reducing SL Entailments to bi-EPs. **Notation** bi-EP$(\phi, \psi) \leftarrow (\theta, S_Z, S_R)$ denotes a judgement, "the problem bi-EP$(\phi, \psi)$ has a solution $(\theta, S_Z, S_R)$".

$$\frac{\text{Axiom}}{\bot \longrightarrow \text{emp}} \qquad \frac{\phi \longrightarrow \text{emp} \quad \psi \longrightarrow \text{emp}}{\phi * \psi \longrightarrow \text{emp}} \qquad \frac{\phi \longrightarrow \text{emp}}{\phi \land P \longrightarrow \text{emp}} \qquad \frac{\phi(\beta) \longrightarrow \text{emp}}{\exists \alpha. \phi(\alpha) \longrightarrow \text{emp}} \text{ for a fresh fixed } \beta \qquad \frac{\text{IdEle}_I(T, D) \quad x \in D}{x \mathbin{\S} T \longrightarrow \text{emp}}$$

$$\frac{\text{Axiom}}{\text{emp} \longrightarrow \top} \qquad \frac{\text{emp} \longrightarrow \phi \quad \text{emp} \longrightarrow \psi}{\text{emp} \longrightarrow \phi * \psi} \qquad \frac{\text{emp} \longrightarrow \phi \quad P}{\text{emp} \longrightarrow \phi \land P} \qquad \frac{\text{emp} \longrightarrow \phi(\beta)}{\text{emp} \longrightarrow \exists \alpha. \phi(\alpha)} \text{ for a fresh free } \beta \qquad \frac{\text{IdEle}_E(T, D) \quad x \in D}{\text{emp} \longrightarrow x \mathbin{\S} T}$$

Fig. 6. Transformations to or from empty, used to solve two subgoals raised by rule (bi-EP) and (bi-EP$_R$).

$$\frac{\text{bi-TP}(T, U, \{(x, z)\}) \leftarrow (\theta, f, Z, R) \quad \text{Trans}(U, \leq)}{\text{bi-EP}(x \mathbin{\S} T, y \mathbin{\S} U) \leftarrow (\theta \land (x, z) \in \text{dom}(f) \land \pi_1(f(x, z)) \leq y \,,\, z \mathbin{\S} Z \,,\, \pi_2(f(x, z)) \mathbin{\S} R)} \text{(biTP)}$$
where $\pi_i$ denotes the $i^{\text{th}}$ projection of a tuple.

$$\frac{\text{Axiom}}{\text{bi-EP}(\text{emp}, \psi) \leftarrow (\text{true}, \psi, \text{emp})} \text{(emp}_L) \qquad \frac{\text{Axiom}}{\text{bi-EP}(\phi, \text{emp}) \leftarrow (\text{true}, \text{emp}, \phi)} \text{(emp}_R)$$

$$\frac{\text{Axiom}}{\text{bi-EP}(\bot, \psi) \leftarrow (\text{true}, \text{emp}, \bot)} (\bot_L) \qquad \frac{\text{Axiom}}{\text{bi-EP}(\phi, \top) \leftarrow (\text{true}, \top, \text{emp})} (\top_R)$$

$$\frac{\text{bi-EP}(\phi_1, \psi) \leftarrow (\theta_1, Z, R_1) \quad \text{bi-EP}(\phi_2, Z) \leftarrow (\theta_2, Z', R_2)}{\text{bi-EP}(\phi_1 * \phi_2, \psi) \leftarrow (\theta_1 \land \theta_2, Z', R_1 * R_2)} (*_L)$$

$$\frac{\text{bi-EP}(\phi, \psi_1) \leftarrow (\theta_1, Z_1, R) \quad \text{bi-EP}(R, \psi_2) \leftarrow (\theta_2, Z_2, R')}{\text{bi-EP}(\phi, \psi_1 * \psi_2) \leftarrow (\theta_1 \land \theta_2, Z_1 * Z_2, R')} (*_R)$$

$$\frac{\text{bi-EP}(\phi, \psi) \leftarrow (\theta, Z, R)}{\text{bi-EP}(\phi \land P, \psi) \leftarrow (P \rightarrow \theta,\ Z,\ R)} \text{(pure-}\land_L) \qquad \frac{\text{bi-EP}(\phi, \psi) \leftarrow (\theta, Z, R)}{\text{bi-EP}(\phi, \psi \land P) \leftarrow (\theta \land P, Z, R)} \text{(pure-}\land_R)$$

$$\frac{\text{bi-EP}(\phi, \psi(\beta)) \leftarrow (\theta, Z, R)}{\text{bi-EP}(\phi, \exists \alpha. \psi(\alpha)) \leftarrow (\theta, Z, R)} (\exists_R) \text{ for a fresh free variable } \beta$$

Fig. 7. Reducing bi-EPs to bi-TPs. Symbols $\phi, \psi, Z, R$ range over S, and $\theta, P$ over FOL formulas.

In the rest of this section, we return to elaborate on the reduction from SL entailments to bi-TPs, which is split into two stages: the construction of bi-EPs as in §8.2.1 and the reduction from bi-EPs to bi-TPs as in §8.2.2.

8.2.1 *Reduction from SL Entailments to bi-EPs.* This reduction starts from judgment $\theta \mid S \vdash G$ that encodes the task of inferring the proof obligation $\theta$ of entailment $(S \rightarrow G)$.

$$(\theta \mid S \vdash G) \triangleq (\text{if } \textit{FOL} \text{ formula } \theta \text{ holds, SL formula } (S \rightarrow G) \text{ holds})$$

Given an SL entailment $S \rightarrow G$, for $S \in S$ and $G \in G$, our reasoner first initiates a goal $(\theta \mid S \vdash G)$, setting $\theta$ as a fresh free variable to be instantiated by the later reasoning process. The obtained instantiation of $\theta$ is the output of our reasoner, which is a proof obligation strong enough to prove the given entailment. This proof obligation is sent to automatic solvers or manual proof works.

Our reasoner then applies rules in Fig. 5 exhaustively in a backward manner to decompose the goal $(\theta \mid S \vdash G)$ into a series of bi-EPs and subgoals in forms emp $\longrightarrow$ S and S $\longrightarrow$ emp. (bi-EP) and (bi-EP$_R$) are the rules that convert entailments into bi-EPs. Other rules eliminate connectives on the right-hand side of entailments, to convert them into a form applicable by (bi-EP) or (bi-EP$_R$).

The obtained bi-EPs are handled in the next subsection. Subgoals in forms emp $\longrightarrow$ S and S $\longrightarrow$ emp are solved by rules in Fig. 6 via backward reasoning. Two algebraic properties of predicates are used here to provide abstract domains $D$ about empty resource. IdEle$_I$ (IdEle$_E$) specifies a domain $D$ of abstractions that can transform to (be made from) empty.

$$\text{IdEle}_I(T, D) \triangleq \forall x \in D.\ (x \mathbin{\S} T \longrightarrow \text{emp}) \qquad \text{IdEle}_E(T, D) \triangleq \forall x \in D.\ (\text{emp} \longrightarrow x \mathbin{\S} T) \qquad (\textbf{IE}_I\ \&\ \textbf{IE}_E).$$

Besides, rules (bi-EP) and (bi-EP$_R$) require that no $\exists$ occurs on an entailment's left-hand side. To eliminate any $\exists$ in this position, we can apply rewriting to move it to the outermost scope (by the rewrite rules used in Skolemization) and then apply (S$_\exists$) to eliminate it.

Additionally, our reasoning process can be extended to support overloading multiple **wp**-rules on one program operation and resolving the proper rule to apply. As we primarily focus on deriving SL entailments using the TP/bi-TP solver, we leave this extension to Appendix G.

*8.2.2 Reductions from bi-EPs to bi-TPs.* Having reduced decision problems of SL entailments to bi-EPs in the previous subsection, now we present the reduction from bi-EPs to bi-TPs.

This reduction is realized by applying the rules in Fig. 7 exhaustively in a backward manner. Rule (biTP) reduces a bi-EP between predicate applications to a bi-TP. All other rules are used to eliminate connectives, ultimately reducing a bi-EP between composite formulas to bi-EPs between predicate applications, into a form to which (biTP) can apply.

In rule (biTP), Trans$(U, \leq)$ is an algebraic property specifying that the order ($\leq$) is a lower approximation to the entailment relation of the abstractions of refinement relation $T$.

$$\text{Trans}(U, \leq) \triangleq \text{for any } x, y \text{ such that } x \leq y, \text{ there is } x \mathbin{\S} U \longrightarrow y \mathbin{\S} U \qquad (\textbf{Tr})$$

To explain the rationale behind (biTP), assume bi-TP$(T, U, \{(x, z)\})$ has an answer $(\theta, f, Z, R)$.

$$(x, z) \mathbin{\S} (T * Z) \xrightarrow{\text{by the bi-TP's answer}} f(x, z) \mathbin{\S} (U * R) \xrightarrow{\text{by property Trans}(U, \leq)} (y, \pi_2(f(x, z))) \mathbin{\S} (U * R)$$

The answer provides us a transformation from $T * Z$ to $U * R$ as illustrated by the first arrow in the above diagram. This transformation also claims a proof obligation $\theta \wedge (x, z) \in \text{dom}(f)$ according to the definitions of TP and bi-TP. The abstraction returned by the transformation is $f(x, y)$ while the goal bi-EP$(x \mathbin{\S} T, y \mathbin{\S} U)$ expects $y$. It forces us to subsequently apply property Trans$(U, \leq)$ to transform $\pi_1(f(x, z))\mathbin{\S}U$ into $y\mathbin{\S}U$ as illustrated by the second arrow. This application yields a proof obligation $\pi_1(f(x, z)) \leq y$. Recall that $\pi_i(x_1, x_2) \triangleq x_i$ is the $i^{\text{th}}$ projection of a tuple. Consequently, the bi-EP$(x \mathbin{\S} T, y \mathbin{\S} U)$ has a solution as that presented in rule (biTP).

## 8.3 Matching Goals and Hypothesis

When multiple hypotheses and goals are involved in an entailment, e.g., $\phi_1 * \cdots * \phi_n \longrightarrow \psi * \cdots$, we initially treat every hypothesis as possibly transformable into part of a goal. Thus, we decompose the entailment to smaller bi-abductive entailments from $(\phi_1, \psi), (\phi_2, \psi), \cdots$, iteratively asking every hypothesis whether it entails any component of $\psi$. For each entailment query, our reasoner wields its full power in applying our TP/bi-TP rules, to try and prove an entailment holds. As an example, a proof tree for deriving $x \mathbin{\S} T * y \mathbin{\S} U \longrightarrow y \mathbin{\S} U * x \mathbin{\S} T$ is presented in Appendix H.

In practice, many obtained entailment subgoals are quickly discarded by falling back to (FB$_2$), for example due to lack of applicatable rules. Most predicates are parameterized by identifiers like memory addresses or domains like initial indexes and lengths of slices. These help syntacically guide the solver in applying relevant transformations. In principle the solver can choose a wrong transformation if multiple options are available, which would cause spurious failure of the reasoning process, but our reasoning rule templates are carefully chosen and given priorities in the reasoning system to minimize overlap (a user must still take care when providing their own ad-hoc rules §6.3). We did not observe erroneous rule application in practice in our case studies (§10).

Also, in principle our search strategy has at least quadratic worst-case complexity, although we observe much better practical performance in our evaluation (§10.2).

### 8.4 Handling Existential Quantification and Evars

Correct instantiation of ∃-quantification and evars is a common problem in SL automation frameworks. We describe our strategy and briefly compare it that of RefinedC and Diaframe.

First, our assertion syntax of $x \mathbin{\text{\textsection}} T_a$ differentiates the predicate argument $x$ and predicate parameter $a$. Predicate parameter $a$ is used to identify the name or the address of a resource, or to indicate the domain of a slice. Since our reasoning process is guided by predicates ($T_a$), only the evars occurring in predicate parameters are influential to the reasoning process. Evars in predicate arguments, *FOL* constraints (e.g., the len(?$x$) = 2 in ?$x \mathbin{\text{\textsection}}$ List ∧len(?$x$) = 2), or any other place, are left uninstantiated until they are presented in the final proof obligations, which are solved by *FOL* solvers like Isabelle's Sledgehammer. These evars are then instantiated by the solvers.

Evars in predicate parameters can occur in side conditions, which then affect the reasoning process. First, we simplify side conditions. Then, if the side condition is an equation, we apply unification *only when* one side of the equation is an evar. It provides more confidence to believe the instantiation is correct. For any other side conditions, the reasoner uses Isabelle's built-in tactic `auto` to instantiate evars, which involves a limited strategy for unsafe instantiation, correctable by manual intervention.

RefinedC [64] employs a hybrid heuristic. First, it seals the created evars wherever possible to prevent premature instantiation due to Coq unification. When an evar occurs a side condition, which is an equation, RefinedC removes the seal and tries to unify the equation's two-hand sides, which can badly instantiate some evar, causing a provable goal to be unprovable without manual intervention. For other forms of side conditions, RefinedC can use a set of user rules to simplify the conditions into a unifiable form. Our system attempts to simplify and unify conditions in roughly equivalent places, although using the built-in capabilities of Isabelle's `auto` or Sledgehammer.

Diaframe [50] includes a particularly strong reasoning system for deferring the introduction of evars, due to their frequent folding and unfolding of invariants. As the authors note [50], ordering problems between the introduction of evars (through elimination of existential quantifiers) and the automated unfolding of predicates (which may introduce new quantifiers with conflicting scopes) can threaten the completeness of automation strategies without backtracking. Our automation simply eliminates existential quantifiers eagerly — RefinedC reports an analogous goal-directed approach [64]. This approach makes particular sense for us since we currently manually annotate all predicate unfolding points — ordering issues can be addressed by moving the position of the existing annotation. We leave further automation of predicate unfolding, and related solutions for handling existential quantifiers, to future work.

## 9 Automatically Proving the Algebraic Properties of Predicates

Section §8 has completed the formalization of our SL reasoner. However, this does not conclude our work. The SL reasoner requires users to provide the algebraic properties of their predicates, which still demands manual effort. To minimize this effort, this section presents algorithms for automatically proving the algebraic properties of predicates, aiming to limit the manual effort to specifying the arguments of the algebraic properties and to proving *FOL* proof obligations.

The algebraic properties involved in the paper — $IE_I$, $IE_E$, **Tr**, **TF**, **SH**, **SA**, **SD**, **S1**, **S0** — are defined based on SL entailments belonging to **E** (cf. §8.1). We name these entailments as their *definitional entailments*. To prove a property, it suffices to prove its definitional entailment. Let us first consider non-recursively defined predicates. The algorithm for proving a property $\mathcal{P}$ of a predicate (operator) $T$ is generically formalized as follows.

**Routine 3** (Proving a property $\mathcal{P}$ for a non-recursively defined predicate (operator) $T$).
(1) Construct the definitional entailment of $\mathcal{P}$, denoted as $\phi \longrightarrow \psi$.
(2) Unfold the definition of $T$ in $\phi, \psi$, resulting in formulas $\phi', \psi'$ comprising connectives and the component predicates used to define $T$.
(3) Send the entailment $\phi' \longrightarrow \psi'$ to our SL algorithm described in §8 and §6, which uses algebraic properties of the component predicates to prove the entailment.

As an instance, and also as an exception requiring specific handling, the algorithm for proving Functor property is formalized as follows.

**Routine 4** (Proving Functor$(F, m, d)$, for a non-recursive predicate $F$). Given $(F, s, z)$, in order to prove Functor$(F, m, d)$, the routine first constructs **TF**'s definitional entailment $x \mathbin{\text{\textsection}} F(T) \longrightarrow m(f)(x) * F(U)$, fixing $x \in \mathrm{dom}(m(f))$. Then $F$ is unfolded in the formula sending the resulting entailment to the SL algorithm described in §8 and §6, *after* prepending the TP rule $\dfrac{\text{Axiom}}{\text{TP}(T,U,D') \leftarrow (D' \subseteq D, f)}$ into the sequence $\Re$ of the rules used by the TP/bi-TP solver Routine 1. The registration of the TP rule is the only difference between the routine and the generic process Routine 3.

Since our SL algorithm accepts only entailments belonging to **E**, it restricts the algorithms above to support only predicates whose definitions are within **S**. For other predicates, their properties must be proven manually. In our case studies, only two predicates have definitions that fall outside **S**: the identity refinement Id axiomatically introduced in §4, and the stepwise refinement operator $(T; U)$ which is also defined in §4. Readers may recall $x \mathbin{\text{\textsection}} \mathrm{Ref}_a(T)$. In our implementation, it is actually defined using the stepwise operator $(T; U)$.

*For recursively defined predicates,* the automatic proving is complicated because it is necessary to use induction in the proofs. We adopt an intuitive but not necessarily complete strategy for induction. To keep the discussion simple, we introduce a lemma stating that any formula in **S** can be equivalently represented into a predicate application form $y \mathbin{\text{\textsection}} U$. It allows us to represent a recursive definition as $x \mathbin{\text{\textsection}} T \triangleq f(x) \mathbin{\text{\textsection}} F(T)$ for some $f, F$. We assume a well-founded recursion mechanism is provided in the underlying proof assistants, allowing this recursive definition. Our implementation uses Alexander's work [41]. We also assume the mechanism generates the well-founded relation $\mathcal{R}$ that orders the arguments in the recursive calls of $T$. The induction rule of $T$ has the following form: For any proposition $P$ about $x \mathbin{\text{\textsection}} T$,

$$\dfrac{P(x \mathbin{\text{\textsection}} T) \text{ holds, if for any } y \text{ such that } y\,\mathcal{R}\,x,\, P(y \mathbin{\text{\textsection}} T) \text{ holds}}{P(x \mathbin{\text{\textsection}} T) \text{ holds for any } x} \text{ (Ind)}$$

The definitional entailments of our algebraic properties take one of two forms:

$$\text{(I).}\ x \mathbin{\text{\textsection}} T \longrightarrow h(x) \mathbin{\text{\textsection}} U \qquad \text{(II).}\ x \mathbin{\text{\textsection}} U \longrightarrow h(x) \mathbin{\text{\textsection}} T$$

for certain $h$ and $U$. Note that $T$ may occur in the expression of $U$. Let us consider form (I) only as the case in form (II) is symmetric.

Initiate a definitional entailment in form (I) as the initial proof goal. Our algorithm first applies rule Ind, instantiating $P$ as $\lambda X.\,(X \longrightarrow h(x) \mathbin{\text{\textsection}} U)$. This results in the following proof state.

$$\frac{\text{Inductive Hypothesis (IH):} \quad \forall a \in pre(x).\ a \mathbin{\mathaccent\cdot\cdot} T \longrightarrow h(a) \mathbin{\mathaccent\cdot\cdot} U}{\text{Proof Goal:} \quad x \mathbin{\mathaccent\cdot\cdot} T \longrightarrow h(x) \mathbin{\mathaccent\cdot\cdot} U}$$

Here, $pre(x) \triangleq \{y \mid y\ \mathcal{R}^+\ x\}$ denotes the set of elements related to $x$ by $R^+$, where $\mathcal{R}^+$ is the strict transitive closure of $\mathcal{R}$. Next, in order to leverage the IH, which applies to the strict antecedents of $x$, our algorithm unfolds $x \mathbin{\mathaccent\cdot\cdot} T$, resulting in,

$$\text{Proof Goal:} \quad f(x) \mathbin{\mathaccent\cdot\cdot} F(T) \longrightarrow h(x) \mathbin{\mathaccent\cdot\cdot} U.$$

Our algorithm requires that $F$ is a functor and to provide property $\text{Functor}(F, m, d)$. The property can be synthesized automatically by using Functor composition, following the process in §2. From this functor property and the IH, our algorithm deduces a lemma $f(x) \mathbin{\mathaccent\cdot\cdot} F(T) \longrightarrow m(h)(f(x)) \mathbin{\mathaccent\cdot\cdot} F(U)$ and produces a proof obligation $d(f(x)) \subseteq pre(x)$. Given this lemma, to show goal (9), it suffices to show that

$$\text{Proof Goal:} \quad m(h)(f(x)) \mathbin{\mathaccent\cdot\cdot} F(U) \longrightarrow h(x) \mathbin{\mathaccent\cdot\cdot} U.$$

Recall that $T$ may occur in the expression of $U$. Because we have unfolded $T$ on the left-hand side once in step 9, the occurrences of T there have a recursion depth one level lower than those in U on the right-hand side. To balance this, our algorithm performs a single-level unfolding for every occurrence of $T$ in $U$ on the right-hand side, thus equalizing the recursion depth on the two sides.

At this point, the algorithm assumes that all reasoning processes about induction have been completed. This assumption, while potentially incorrect, allows the algorithm to proceed; if it is not true, the algorithm is still sound but incomplete. Given this assumption, our algorithm then passes the proof goal to the SL entailment algorithm formalized in §8 and §6, following the same process for non-recursive predicates.

## 10 Evaluation and Case Studies

In this section, we perform an evaluation to: (1) Validate the feasibility of the reasoner in verifying real data structures and algorithms written in an imperative language having a block memory model that resembles CompCert. (2) Demonstrate that our approach indeed reduces the need for manually crafting predicates' reasoning rules (in terms of refinement-type system, manual typing rules). (3) Evaluate the degree of automation of the generic reasoner. We show that the result is similar to the state-of-the-art works. (4) Demonstrate the effectiveness of our algorithms for proving the algebraic properties. The algorithms automate the verification condition generation for all properties in the case studies in this section and ~95% of the properties used in our system implementation.

### 10.1 Implementation Details: Semantic Formalization and Fictional Separation

To perform the evaluation, we formalize our theory in Isabelle/HOL [53], implement the automation algorithms based on a semantics formalized in Isabelle/HOL, and evaluate the reasoner on 592 lines of programs involving 10 different data structures, from usual structures like linked lists to more challenging ones, such as AVL trees and bucket hash maps. As a side note, our implementation is based on an *sp*-transformer for the sake of symbolic execution, which is essentially equivalent to the *wp*-transformer.

**The semantics** of our generic heap language supports pointer arithmetic, the address-of operator for memory objects, and fixed-size integers. The memory model is based on a map (block × offset → byte) from memory blocks and offsets to the base addresses of the blocks to bytes. A pointer is represented as a pair (block × offset) (this resembles the basic CompCert memory model [45]). Using a pointer (*blk*, *ofs*) to access a location beyond the boundaries of the block *blk* of the pointer always fails, and pointer comparison between pointers of different memory blocks always returns false. These simplifying assumptions mean that we do not grapple with complex questions of pointer provenance that are raised by program logics over a fully-faithful C semantics with undefined behaviour [64].

Table 2. Property derivation of selected predicate operators.

| Operator(s) | Abst. | Tr | IE | TF | SH | SD/SA/S1 | $\mathcal{R}$ | $\mathcal{R}'$ | $\mathcal{M}$ | $\mathcal{M}'$ |
|---|---|---|---|---|---|---|---|---|---|---|
| Record | tuple | 0 | 0 | 0 | 0 | × / 0 / 0 | 20 | 25 | 0 | 0 |
| Variable, Ref | identity | 0 | 0 | 0 | 0 | 0 / 0 / 0 | 6,6 | 0,2 | 0 | 0,4 |
| Quantifier ✶ | map | 0 | 0 | 0 | 0 | 0 / 0 / 0 | 19 | 38 | 2 | 0 |
| Array Slice | list | 0 | 0 | 3+1 | 0 | 0 / × / 0 | 25 | 4 | 4 | 0 |
| Linked List | list | 0 | 0 | 0 | × | × | 9 | 0 | 0 | 0 |
| Dynamic Array | list | 3+1 | × | 0 | × | × | 7 | 4 | 0 | 0 |
| Binary Tree | tree | 0 | 0 | 1+1 | × | × | 10 | 8 | 0 | 0 |
| Search Tree | map | 3+3 | 0 | 2+1 | × | × | 5 | 0 | 0 | 0 |
| Lookup AVL | map | 3+3 | 0 | 0 | × | × | 5 | 0 | 0 | 0 |
| Bucket Hash | map | 2+8 | × | 14+9 | × | × | 7 | 0 | 0 | 0 |

Table 3. Evaluation of our verification framework over the cases.

| Group (#) | $\mathcal{M}/\mathcal{M}'$ | $\mathcal{R}/\mathcal{R}'$ | Prop | Anot | Fold | Othr | Prf | Aux | Ovh | Ovh* | LoC | Time |
|---|---|---|---|---|---|---|---|---|---|---|---|---|
| Rational (4) | 0/0 | 20/358 | 0.02 | 0.37 | 0.37 | 0 | 0 | 0 | 0.40 | - | 43 | 0.4s + 2.9min + 3s |
| Link-List (10) | 0/0 | 48/1120 | 0.04 | 0.19 | 0.19 | 0 | 0 | 0 | 0.24 | 0.25 [64] | 67 | 0.3s + 0.7min + 8s |
| Quicksort (1) | 0/0 | 57/750 | 0 | 0.39 | 0 | 0 | 0.33 | 0 | 0.72 | 4.95 [70] | 18 | 0.5s + 3.4min + 50s |
| Bin. Search (2) | 0/0 | 32/706 | 0 | 0.24 | 0 | 0 | 0 | 0 | 0.24 | 0.60 [64] | 33 | 0.3s + 1.0min + 12s |
| Dyn. Array (8) | 0/0 | 87/3838 | 0.05 | 0.19 | 0.18 | 0 | 0 | 0 | 0.24 | 0.19 [65] | 62 | 2.5s + 3.1min + 53s |
| Matrix (6) | 0/0 | 56/1200 | 0.05 | 0.36 | 0.15 | 0 | 0.03 | 0 | 0.44 | 2.86 [69] | 39 | 0.6s + 0.6min + 20s |
| Strassen Al. (2) | 2/8 | 68/4168 | 0.03 | 0.26 | 0.09 | 0.17 | 0.43 | 0 | 0.72 | - | 65 | 2.4s + 3.6min + 47s |
| Binary Tree (3) | 0/0 | 47/3174 | 0.07 | 0.41 | 0.41 | 0 | 0.03 | 0.84 | 1.34 | 1.07 [64] | 46 | 0.7s + 3.6min + 22s |
| AVL Tree (3) | 0/0 | 60/20533 | 0.06 | 0.27 | 0.27 | 0 | 0.11 |  | 1.28 | - | 106 | 6.8s + 6.2min + 201s |
| Buck. Hash (11) | 0/0 | 102/3747 | 0.08 | 0.31 | 0.09 | 0.04 | 0.12 | 0 | 0.51 | 2.44 [13] | 113 | 3.7s + 6.3min + 23s |

We also do not support concurrency [16, 38], address-of operator for local variables, goto, loose expression evaluation ordering [22], and first-class function pointers. However, we believe a low-level block memory model is still sufficient to examine the degree of automation for an SL reasoner.

**Fictional separation** is also supported by our system in basic form. This allows us to write assertions with separating conjunction over an abstract *fictional* memory of records with fields, and relate them to assertions about the underlying concrete memory layout. We follow Jensen and Birkedal's Fictional Separation Logic [35], which is designed for sequential reasoning and perfectly matches our needs. It introduces two PCMs, one for abstract representations (called *fictions*), and another for concrete resources. Denote the PCMs by $\mathcal{A}$ and $C$. A usual assertion logic of SL (also known as Bunched Implications) is built over the fictional PCM $\mathcal{A}$, i.e., $\mathcal{A}$ is the model of the assertion logic, and its assertions are about fictions. So-called *fictional interpretation* $I : \mathcal{A} \to 2^C$ is a map from fictions to sets of concrete representations. It converts assertions about fictions into assertions about concrete resources. For further details, we refer readers to their work [35]. In our work, the concrete PCM $C$ encompasses the concrete memory model, and the fictional PCM $\mathcal{A}$ encompasses an abstract memory model (logical-address → permission × value) where we associate logical addresses with permissions and high-level value representations. The logical-address ≜ (block × field-name list) uses a path of field names to locate the address of a field entry. The value is an algebraic data type that deep-embeds integers, arrays, structures, and their combinations. The permission ≜ (positive rational) annotates the ownership of each logical address. Permission modality $\oplus$ is implemented based on it, $(w \models x \,\S\, n \oplus T) = (w' \models x \,\S\, T)$, where $w'(a) = (p, v)$ iff $w(a) = (np, v)$. The fictional interpretation $I$ involves 1) how we translate logical addresses to physical addresses and how to represent values as bytes; 2) constraining the sum of permissions of every reference to equal 1. As we discuss in §12, we aim to extend our logic with further fictions in future work.

### 10.2 Evaluation

**The property derivation** for the 10 data structures and a modality (the multiplicative quantifier ✻) is detailed in Table 2. The verification condition generation for all the properties is automatic. Users only need to specify the expression of the properties and to prove any Isabelle/HOL proof obligations failed to be proven by ATPs.

In Table 2, *Abst* denotes the abstract representations used in the refinements of the data structures. (×) indicates that the property does not hold for the operator. $n+m$ denotes that the derivation generates pure proof obligations that are solved automatically using $n$ tactic hints and $m$ tactic arguments. If the proof is fully automated without any hints or arguments needed, we write 0. $\mathcal{R}$ denotes the number of rules that are instantiated from algebraic properties and are used at least once in verifying programs. $\mathcal{R}'$ denotes the number of such rules used in deriving other properties. $\mathcal{M}$ denotes the number of handcrafted rules. $\mathcal{M}'$ denotes the number of lines of manual configurations required for the derived rules.

**The verification for the programs** is detailed in Table 3. Specifications of the programs entail refinements relating the programs to abstract representations. For example, the insert operations in the lookup AVL and the bucket hash are related to the same abstraction, update of partial maps,

$$\{f \,\S\, \text{AVL}_{ptr}(T) * k \,\S\, \text{val}(\mathbb{N}) * x \,\S\, \text{val}(T)\} \, \text{insert\_AVL}(ptr, k, v) \, \{f(k \mapsto x) \,\S\, \text{AVL}_{ptr}(T)\}$$
$$\{f \,\S\, \text{Hash}_{ptr}(T) * k \,\S\, \text{val}(\mathbb{N}) * x \,\S\, \text{val}(T)\} \, \text{insert\_Hash}(ptr, k, v) \, \{f(k \mapsto x) \,\S\, \text{Hash}_{ptr}(T)\}.$$

These specifications cover partial correctness only. The support for total correctness is left to future works as it demands a predicate property for order-preserving.

In Table 3, routines are grouped, and their number in each group is labeled in parentheses. Column *LoC* shows the number of lines in the source code. Most other metrics are based on the number of lines and presented as ratios relative to LoC. These include: *Prop*, which represents manual specifications and proofs for the algebraic properties *required* by the cases (subtyping rules of the dynamic array and hash table are not required in verifying their operations); *Anot*, covering all annotations; *Fold*, annotations specifically for folding and unfolding predicates; *Othr*, other annotations excluding Unfold and loop invariants. *Prf*, manual proofs for proof obligations; *Aux*, auxiliary Isabelle/HOL theories for the abstract representations used in the refinements. If Aux = 0, the abstract representations come from Isabelle system libraries. *Ovh* = Property + Annot + Prf + Aux, measuring the ratio of overhead to LoC. $Ovh^*$ is the best Ovh for the same kind of data structures and algorithms (but not the same implementation) found in the literature of SL-based verification for real languages, with citations in brackets. (-) indicates that the case or its statistics are not observed in the literature or released sources, to the best of our knowledge.

*Time* given in format $a + b + c$ denotes the time taken to verify the programs from scratch. Portion $a$ is the time taken in applying the SL rules in the reasoner (§6 and §8), while portions $b+c$ concern the time taken to discharge proof obligations arising during the reasoning process. Portion $b$ represents proof search - first with basic tactics like auto, then with Isabelle's Sledgehammer if otherwise unsuccessful. Once a proof script is found by this search, it is cached for future use - the invocation of Sledgehammer and subsequent caching are fully automated by our framework. Portion $c$ represents the time taken to verify a proof script originally discovered in portion $b$.

Because of our caching, replaying the same verification (e.g. validating a previous result) takes only time $a + c$. Times are measured on an Intel i9-13900K CPU (16 cores) with 32 GB memory.

**Annotations.** Among the annotations used in the case studies, 64% are used for folding and unfolding predicates, which constitute the largest portion as shown in column Unfold, and 25% are used for loop invariants, while only 11% are used to guide the reasoning process. Different from existing methods equipped with automatic mechanisms to (un)fold definitions, our reasoner never silently unfolds predicates and requires any (un)folding to be explicitly annotated. Fortunately, writing these (un)folding annotations is not as challenging as providing reasoning guidance annotations. Instead, it can be more straightforward because folding and unfolding are usually used in pairs when the reasoning involves the internal implementations of data structure operations. Although our reasoner preserves predicate abstractions, unfolding is unavoidable to verify internal implementations; otherwise, there is no way to verify them.

**Generation of reasoning rules.** Certainly, any individual case study can be automated using a tailored heuristic and hand-crafted reasoning rules. However, *the key distinction of our work* lies in its generality: all the case studies are automated by *one* generic algorithm, using rules automatically instantiated from *one* theory about algebras of refinement transformations and relying on barely *no* manually-written ad-hoc rules for predicates, except for those directly handling primitive types like fix-size integers.

Including the entire system implementation, our SL reasoner incorporates ∼300 rules for transforming predicates (which correspond to typing rules in refinement type systems). Of these, ∼80% are generated automatically, while remains are ad-hoc rules for specific predicates, which are typically predicate counterparts of connectives such as ∨, ∧, ∗, and if-then-else.

Regarding the case studies listed in this section, column $\mathcal{R}$ presents the number of distinct instantiated rules used in the verification; $\mathcal{R}'$ indicates the total times of their applications. Columns $\mathcal{M}$ and $\mathcal{M}'$ represent the numbers of manually written rules and annotations to apply them, respectively. This suggests our method almost eliminates the need for manual rules in the case studies. The two manual rules are used for splitting and merging matrix into and from four sub-matrixes in Strassen's algorithm. The rules are manually specified but automatically derived, after unfolding the matrix abstractions into two-dimensional arrays. The eight manual applications are for applying them. The rules are not automated because (1) the scalars of matrixes are two-dimensional intervals whose addition is not associative, and (2) our reasoner does not support slicing for non-associative scalar addition.

**Non-trivial transformations** by generated rules are essential in the case studies.

For instance, let's consider the access to a field of a record in memory. Recall that $x \, \S \, \{a: T\}$ specifies a record having a field $a$ and the field has a value refining $x$ w.r.t. $T$; we use $\{a: T\} \ast \{b: U\}$, abbreviated as $\{a: T, b: U\}$ to represent a record of two fields; $a.b$ is a path (§ 5.3) to a field in a nested record. From the definitions of Ref and val in Fig. 4 we specify a load operation as $\{x \, \S \, \mathrm{Ref}_{addr}\{a: T\}\}\mathtt{load}(addr.a)\{v. \; x \, \S \, \mathrm{Ref}_{addr}\{a: T\} \ast x \, \S \, \mathrm{val}_v \, T\}$. Given a program state specified by $((x, y), z) \, \S \, \mathrm{Ref}_{addr}\{a: \{b: T_1, c: T_2\}, d: T_3\}$, the load operation of the field $a.b$ requires to transform the state specification to $(1. \; x \, \S \, \mathrm{Ref}_{addr}\{a.b: T_1\}) \, \ast \, (2. \; (y, z) \, \S \, \mathrm{Ref}_{addr}\{a: \{c: T_2\}, d: T_3\})$, so that we can apply the triple of the load by using the frame rule to frame out (2). This transformation requires properties **SH** of Ref, **SH** of $\lambda T. \{a: T\}$, and also **SA** of $\lambda T. \{a: T\}$ that allows $\{a: \{b: T\}\} = \{a.b: T\}$.

**SD** is necessary in the case study of Quicksort. As a divide-and-conquer algorithm, a key step of Quicksort is to divide (the abstraction of) an array slice into two and send them respectively to recursive calls. The divide and merge of the slices require the **SD** property of array slices (§ 5.3).

As our third instance, Strassen's matrix multiplication is another divide-and-conquer algorithm. The partitioning of matrix slices in this algorithm requires **SD** and **SH**, as discussed in § 6.6.

The fourth instance is the use of scalar unit property **S1** in Bucket Hash. A bucket hash contains a sequence of buckets, and each bucket is represented by a dynamic array. In the case study, we use finite separating quantifier $\{x_i\}_{i \in I} \mathbin{\text{\textfmty}} \divideontimes_{i \in I} T_i \triangleq (x_1 \mathbin{\text{\textfmty}} T_1 * \cdots x_n \mathbin{\text{\textfmty}} T_n)$ for $I = \{1, \cdots, n\}$, to collet the sequence of the buckets. To access the $k^{\text{th}}$ bucket represented as $x_k \mathbin{\text{\textfmty}} T_k$, it requires a transformation $\{x_i\}_{i \in I} \mathbin{\text{\textfmty}} \divideontimes_{i \in I} T_i \longrightarrow \{x_i\}_{i \in I \setminus \{k\}} \mathbin{\text{\textfmty}} \divideontimes_{i \in I \setminus \{k\}} T_i * x_k \mathbin{\text{\textfmty}} T_k$. This transformation is derived from the properties **S1** and **SD** of $\divideontimes$.

**The efficiency of our algorithm** is estimated by correlating the number of rule applications against the number of connectives and predicate operators in reasoning every step of program operation (Fig. 8). We use the number of rule applications because it is a fair indicator independent of specific implementations to measure the time cost of the reasoning process. The time for a single rule application is generally constant and takes <0.15ms in our implementation. Although the theoretical time complexity is at least quadratic due to the bi-abduction process (§8), the statistics show the actual efficiency is much better than the theoretical worst-case quadratic time. In practice, our method is fast because predicate operators form expression trees. The reasoning process is segmented by tree hierarchies, with each segment focusing solely on the children of each tree node. When the number of children (i.e., components in a hierarchy, such as fields in a record) remains constant relative to the scale of the entailment formula, a linear result emerges.

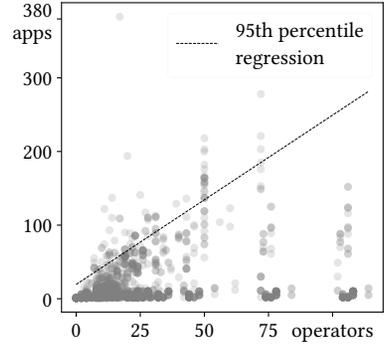

Fig. 8. Each point represents a bi-EP instance. The X-axis is the number of type operators and connectives in this bi-EP; the Y-axis is the number of applications of rules used to solve the bi-EP.

**The degree of automation.** Before any discussion, we must acknowledge that the degree of automation in foundational verification cannot be simply measured by lines of code or annotations. Considerable differences exist in semantic formalizations and the automation capabilities provided by underlying proof assistants. Moreover, the manual effort for debugging the reasoning and providing guidance annotations is more arduous than adding (un)folding-annotations that are generally placed at the beginning or end of internal operations. Although quantitatively measuring the degree of automation in foundational verification is challenging, columns Ovh, Ovh* suffice to show that the instantiated automation from our generic reasoner is at least comparable with the state-of-the-art works based on specially designed reasoning rules and heuristics. Importantly, our automation stems from a general theory with automatically instantiated rules, conferring better generality than existing methods.

### 10.3 Qualitative Comparison

One weakness of our tool in comparison to related works is our reliance on explicit predicate (un)folding annotations. Regarding the case studies discussed above, recent works, particularly RefinedC and Quiver, have achieved a high degree of automation here, with almost no annotation (except loop invariants) required.

For example, the maintenance routine in AVL involves 4 branches to determine which subtrees should be rotated. When different subtrees are rotated, different predicates should be unfolded, which in our tool requires unfolding and folding notations in each branch. Worse, the routine has multiple return statements, each one in a branch, causing a predicate unfolded by one annotation to require multiple annotations to fold it back. Consequently, the routine requires 5 unfolding annotations and 11 folding annotations. Our unfolding annotations are not smart enough to case-split algebraic data types, causing 4 additional lines to be required to specify the proper shape of the abstract tree representations. Finally, 20 lines are used for the annotations.

As we discuss in §8.4, automatic predicate unfolding strategies of other tools present additional complications for the syntax-directed handling of ∃-quantification and evars. We believe that our system can be extended with more ambitious automatic predicate unfolding, although this would mean that these related challenges would also need to addressed.

Regarding loop invariants, all discussed tools require some manual annotation. In particular, both our work and RefinedC require users to specify the invariant refinement relations (refinement types) of variable states and constraints about the abstract representations (the values of the types).

While our work is focussed on the automatic derivation of predicate transformations wherever possible, our tool is not able to automatically derive the predicate transformations necessary to split and merge the matrix slices of the Strassen's algorithm example (see § 10.2). Like existing tools, we must instead prove and apply the relevant predicate transformations ourselves — in our case requiring 10 additional lines of proof annotation.

Except for these, all the remaining annotations are used to manually prove *FOL* side conditions that cannot be automated. For a side condition, our reasoner first tries to use Isabelle's built-in tactic auto to automatically prove or disprove it within a time limit. If the solver neither proves nor disproves it within the time limit, it will be printed on the screen, and our reasoner proceeds by assuming it is false. If it is actually true, the reasoning can fail. In this case, users must manually prove the side condition and register it with our system, so that the reasoner can apply the proper rule guarded by the condition. Manual effort is also required to prove side conditions in RefinedC. Users need to annotate tactics that augment the default solver to prove the side conditions that fail to be automated. Both our and their solvers can be improved to reduce these these annotations.

The SL used in our work also has expressiveness deficiencies compared with tools [23, 50, 64, 65] built on rich SLs like Iris. Our logic does not support higher-order fiction, higher-order ghost states, and the later modality. It would be interesting to investigate how to apply our algebraic system to later modality, ghost states, update modality, and other connectives specific to these rich logics.

We highlight earlier sections where we make relevant qualitative comparisons to related tools: handling of ∃-quantification and evars (§ 8.4); and discrepancies between the semantics such as our lack of concurrency or provenance support (§ 10.1). See our discussions in Related Work (§ 11) on bi-abduction uses and expressivity of the logic in comparison to other tools.

## 11 Related Work

Regarding *SL-based foundational verification*, the literature contains a wealth of related work. Based on these works, our main advancement is an algebraic approach to generating reasoning rules of non-trivial transformations. The generation is generic for any predicates satisfying certain algebraic properties that can also be automatically proven by our method.

**RefinedC** [64] is a foundational verifier for a major fragment of the C language. It provides an impressive degree of automation comparable to non-foundational tools while being based on an expressive separation logic, Iris [37]. Two key components contribute to its automation capability.

One component is the separation logic programming engine, Lithium, which provides a non-backtracking reasoning mechanism for deriving SL entailments. Lithium has significantly influenced our work. The syntax of our reasoner (§ 8.1) is essentially borrowed from Lithium's syntax. Our reasoning process (Fig. 5) for eliminating connectives mirrors its goal-directed search.

Another component is its refinement type system, which effectively handles low-level programming idioms. A rich concept of subtyping is an important part of this system, allowing various refinement types to be applied to the same data structure. The transformations between predicates studied in our work correspond to their subtypings. Their subtyping rules are given in a $wp$-transformer form, which we simplify into implications between predicate applications. This simplification enables us to develop the algebras of transformations between refinements, forming the basis of our algebraic method for rule generation.

While rules for user-defined types derived by RefinedC are (un)folding rules, our approach furthermore automatically generates subtyping rules and other non-trivial transformation rules for both user-defined and system predicates. These instantiated rules are powerful enough to constitute the core of our SL reasoner. The reasoner demonstrates a promising level of automation, which in many cases is similar to that of RefinedC's reasoner, despite RefinedC's use of manually crafted typing rules. However, RefinedC's memory model involves more complicated matters like provenance, padding of structures, and pointer alignment constraints. It would be interesting to investigate how our rule generation system benefits manually proven rules in RefinedC.

**Diaframe** [50] is a more recent tool for automating foundational verification on top of Iris. Both Diaframe and RefinedC are based on ideas from linear logic programming and share some common methods for reasoning about certain logic connectives. Similarly to RefinedC's typing rules, Diaframe's automation power depends on an extensible set of *hints*. A Diaframe hint is essentially a reasoning rule that plays a similar role to our bi-TP rules, and also uses a similar bi-abduction technique for a similar purpose — answering which part of the source assertion can be transformed to which part of the target assertion, and what is left and missing after this transformation. To extend Diaframe's automation capability on new predicates, new hints must be crafted and proven manually. This is where our generic theory of rule generation could be potentially beneficial. Where Diaframe relies on manually crafted hints, our theory could potentially help it to identify useful hints and generate hints automatically. On the other hand, Diaframe additionally automates Iris-specific modalities and ghost state updates, which we have not considered yet, and many of its discussed examples are on concurrent programs, which we currently do not handle. In future work, we hope to investigate the composition of our techniques with theirs.

**Quiver** [65] is a recent foundational verifier for C that aims to reduce specification overhead by automatically inferring functional correctness specifications. It is also based on the separation logic Iris and the same semantic formalization Caesium [64] with RefinedC. Quiver extends RefinedC's approach in two key ways: it enhances the refinement type system to work under incomplete information about the proof context, and it introduces bi-abduction to infer complete specifications from partial specification sketches provided by users. While Quiver uses bi-abduction for specification inference, our work employs bi-abduction for a different purpose. We use it as part of our reasoning process to decompose entailments between composite formulas into those between predicate applications, which are then reduced to transformation problems. This allows us to apply our reasoning rules generated from algebraic properties, to the transformation problems.

**RefinedRust** [23], built on Iris logic, is a recent approach for verifying both safe and unsafe Rust programs. RefinedRust adapts and enhances RefinedC's refinement type system to accommodate Rust-specific concepts such as borrowing, lifetimes, and "places". Key innovations include borrow names and place types, which enable Rust-specific reasoning about borrowed places.

**Islaris** [63] is an Iris-based framework for verifying machine code against comprehensive and authoritative instruction set architecture specifications for Armv8-A and RISC-V. It adapts RefinedC's Lithium to a logic tailored for tracing instruction's register and memory accesses.

**VST** [1, 12] is an SL-based interactive tool for verifying CompCert [45] C programs. Its automation system is based on a symbolic execution aided by user annotations for intermediate assertions. VST-A [75] extends VST by allowing users to write proofs in annotations directly.

**Bedrock** [13–15, 46] is the initial work for SL-based foundational verification. Bedrock and VST families encode specifications in plain separation logic without systematic use of refinement techniques. Their automation depends on pre-defined heuristics and custom user tactics. In contrast, we aim to propose a systematic theory for automating SL for almost any data structure.

Beyond the works directly related to our approach, other foundational verification works based on SL also make significant contributions in their respective domains [25–28, 36, 39, 47, 48, 58]. Additionally, there is also a wealth of *non-foundational tools* built on top of expressive SLs.

**CN** [59] is a deductive verifier for C language aiming to bridge the gap between verification techniques and real software development. It is based on a carefully designed refinement-type-integrated SL where the produced logical constraints always fall into an SMT fragment known to be decidable. More remarkably, it is built upon an accurate ISO C semantics validated on substantial C test suites. This shows the feasibility of verifying a large fragment of ISO C.

**VeriFast** [33] is an automated, separation logic-based functional correctness verifier for C and Java. Its automation is also based on predefined or user-provided heuristics [71]. It does not provide a rule-generation mechanism for non-trivial transformations like ours.

Finally, the literature also abounds in automatic solvers for *specific fragments of separation logic*. This line of research began with symbolic heap [2], a decidable fragment of separation logic. Subsequent works extend it by incorporating inductive predicates [20, 21, 43, 44, 67] and arrays [7], introducing various automation techniques including superposition calculus [51], lemma synthesis [34], model checking [8, 42], symbolic execution [3, 55], and SMT solvers [52, 56]. Initially, these approaches focus on shape analysis for memory safety [4, 6, 9, 10, 34]. Subsequent works extend to functional correctness on intermediate verification languages [32, 57, 60]. While these approaches have made significant strides in automating reasoning for specific SL fragments, foundational verification often requires more expressive SLs, which is where our work aims to contribute.

## 12 Limitations and Future Work

Many refinements listed in Table 2 do not support **SH**. As explained in the end of § 5.2, this is because the refinements are defined in a direct and simple manner, while the data structures contain control structures that cannot be split and shared between separated abstractions under these simple definitions. As an example of a consequence, we cannot easily verify concurrent programs where portions of a hash map are owned by different threads. We plan to introduce a modality based on *fiction of disjointness* [19, 35] in order to wrap any predicate into a separating-homomorphic one. Similarly, many refinements in Table 2 do not support **SD**. We plan to develop a modality for slicing any predicate, e.g., to slice the mapping abstraction of a hash table into sub-mappings, so that different program modules can own and modify different sub-mappings.

Based on other algebraic already defined, we plan to add the necessary automation for deriving implications like $(l_1 \, \S \, \mathrm{List}_{addr1} \, T) * (l_2 \, \S \, \mathrm{List}_{addr2} \, U) \longrightarrow addr1 \neq addr2$, useful for pointer arithmetic.

Besides, an automatic mechanism for (un)folding predicates definitions, and a mechanism for inferring specifications of algebraic properties are also left to our future works.

## Acknowledgements

This research is partially supported by the project MOE-T1-1/2022-43 (funded by Singapore's Ministry of Education), EPSRC grant EP/Z000580/1. Conrad Watt and Qiyuan Xu are supported by an NTU Nanyang Assistant Professorship Start-Up Grant. The early stage of this research was sponsored by Hangzhou Yunphant Network Technology Co. LTD. We are also grateful to the anonymous reviewers for their valuable comments. The artifact of this work is available in [72, 73].

## A  Complete Formalization of the Assertion Language of our Separation Logic

Section §4 has presented a simplified formalization that captures the novelty of how our separation logic integrates data refinements. In order to be self-contained, the complete formalization of the assertion language is presented in this appendix for interested readers to refer to.

The assertion language of the SL is parameterized by a finite set **P** of SL predicates, and a first-order logic *FOL* with equality. Let $w, x, y, z, t$ range over terms in *FOL*, and $\alpha, \beta$ over variables in *FOL*. Fix a set **P** of symbols to denote SL predicates, and let $T, U$ range over them. The assertion language **F** of our SL includes all standard connectives plus a satisfaction operator ($\Vdash$) borrowed from Hybrid Logic [5, 30] (originally denoted by @).

$$\mathbf{F} \ni \phi, \psi ::= \top \mid \bot \mid \mathsf{emp} \mid T(x) \mid \neg\phi \mid \phi * \psi \mid \phi \wedge \psi \mid \phi \vee \psi \mid \phi \mathrel{-\!\!*} \psi \mid \phi \rightarrow \psi \mid \exists \alpha.\, \phi \mid \forall \alpha.\, \phi \mid t \Vdash \phi$$
$$\mid \text{any other formula in } \textit{FOL}, \text{ e.g., } x = y.$$

Fix a domain of discourse $O$ and an interpretation function $[\![-]\!]$ from *FOL* terms to $O$. Fix a partial commutative monoid $\mathcal{A} = (S, \bullet, \epsilon)$ which we call *Separation Algebra*. Elements in $S$ are called *worlds* and ranged over by $w$.

Additionally, fix an interpretation function $[\![-]\!]' : \mathbf{P} \rightarrow 2^{O \times S}$ that assigns every SL predicate (symbol) a subset of the product of the domain of discourse and the set of worlds.

The semantics of formulas in **F** is defined by forcing relation ($\models$), a binary relation between $S$ and **F**. For $w \in S$, and $\phi, \psi \in \mathbf{F}$,

- $w \models \top$ holds anytime;
- $w \models \bot$ never holds;
- $w \models \mathsf{emp}$ holds iff $w$ is the identity element $\epsilon$;
- $w \models T(x)$ holds iff $(x, w) \in [\![T]\!]'$;
- $w \models \neg\phi$ holds iff $w \models \phi$ does not hold;
- $w \models \phi * \psi$ holds iff there exists $w_1, w_2$ such that $w = w_1 \bullet w_2$ and both $w_1 \models \phi$, $w_2 \models \psi$ hold;
- $w \models \phi \wedge \psi$ iff both $w \models \phi$ and $w \models \psi$ hold;
- $w \models \phi \vee \psi$ iff either $w \models \phi$ or $w \models \psi$ holds;
- $w \models \phi \mathrel{-\!\!*} \psi$ holds iff for any $w'$ such that $w' \models \phi$ holds and $w' \bullet w$ is defined, $w' \bullet w \models \psi$ holds;
- $w \models \phi \rightarrow \psi$ holds iff, given that $w \models \phi$ holds, $w \models \psi$ holds;
- $w \models (\exists \alpha.\, \phi)$ holds iff there is an *FOL* term $t$ such that $w \models \phi[t/\alpha]$ holds;
- $w \models (\forall \alpha.\, \phi)$ holds iff $w \models \phi[t/\alpha]$ holds for any *FOL* term $t$;
- $w \models (t \Vdash \phi)$ holds iff $[\![t]\!] \models \phi$ holds;
- if $\phi$ is an *FOL* formula, $w \models \phi$ holds iff $\phi$ holds according to the semantics of *FOL*.

Notation $\phi[t/\alpha]$ denotes the formula obtained by substituting term $t$ for any occurrences of $\alpha$.

## B  A Many-Sorted Variant of the Assertion Language

As mentioned in §4, we require the worlds of SL assertions to encompass all concrete representations like program states, memory and values, and also all abstract objects that are used as intermediate representations of stepwise refinements, e.g., the $x$ in $y \mathbin{\S} (T; U) \triangleq \exists x.\, x \mathbin{\S} T \wedge (x \models y \mathbin{\S} U)$.

Generally, we want different kinds of objects to be organized in different PCMs, which eases how to define their group operations. The many-sorted variant presented in this section realizes a way for this purpose.

Fix a set $\mathcal{K}$ to denote (the symbols of) all sorts. Every world $w$ is classified into a unique sort, written $\mathbf{k}(w)$. Syntactically, every formula is represented as a tuple of its unsorted expression (an element in **F**) and its sort. Overload $\mathbf{k}(\phi)$ to also denote the sort of formula $\phi$. Every predicate symbol $P$ is also assigned with a unique sort, also denoted as $\mathbf{k}(P)$. A formula is *well-formed formula* (written *wff*) iff the sorts of all subformulas are compatible. Formally,

$$\frac{\phi \text{ is wff} \quad \psi \text{ is wff} \quad \mathbf{k}(\phi) = \mathbf{k}(\psi) = \mathbf{k}(\phi \star \psi)}{(\phi \star \psi) \text{ is wff}} \text{ for } \star = *, \wedge, \vee, -\!*, \rightarrow$$

$$\frac{\phi \text{ is wff} \quad \mathbf{k}(\phi) = \mathbf{k}(Q\alpha.\,\phi)}{(Q\alpha.\,\phi) \text{ is wff}} \text{ for } \star = \exists, \forall \qquad \frac{\phi \text{ is wff} \quad \mathbf{k}(\phi) = \mathbf{k}([\![t]\!]) = \mathbf{k}(t \Vdash \phi)}{(t \Vdash \phi) \text{ is wff}}$$

$$\frac{\phi \text{ is wff} \quad \mathbf{k}(\phi) = \mathbf{k}(\neg\phi)}{(\neg\phi) \text{ is wff}} \qquad \frac{\mathbf{k}(P(x)) = \mathbf{k}(P)}{P(x) \text{ is wff}} \qquad \begin{array}{c} \top \text{ is wff}, \quad \bot \text{ is wff} \\ \text{emp is wff} \end{array}$$

Note that a formula is a tuple of its unsorted expression and its sort. The above rules rule out any such tuples of illegal sorts, ensuring any well-formed formula is a tuple that has a correct sort.

We only consider well-formed formulas henceforth. Thus, *we call a well-formed formula simply a formula*.

The semantics of this many-sorted SL is also many-sorted. Instead of a single huge PCM, the semantics depends on a family of PCMs $\{\mathcal{A}_k\}_{k \in \mathcal{K}}$ indexed by sorts. The semantics is given by the forcing relation ($\models$) between sorted well-formed formulas and the PCMs. Let $\mathrm{expr}(\phi)$ denote the unsorted expression of $\phi$. Note $\mathrm{expr}(\phi) \in \mathbf{F}$. Let ($\models'_{\mathcal{A}}$) denote the old forcing relation defined in Appendix A, which uses PCM $\mathcal{A}$ as its model. The forcing relation ($\models$) in our many-sorted variant is defined based on what we have defined in Appendix A,

$$w \models \phi \text{ iff } \phi \text{ is wff and } \mathbf{k}(w) = \mathbf{k}(\phi) \text{ and } w \models'_{\mathbf{k}(w)} \mathrm{expr}(\phi).$$

The reasoning system can also be migrated accordingly to a many-sorted variant. Our Isabelle implementation is indeed a many-sorted version. We omit the further discussion about the many-sorted reasoning system and refer readers to our source.

As a benefit of this many-sorted SL, the implementation can enjoy shallow embedding and implements PCMs using typeclasses constraints given by the underlying proof assistants. In our Isabelle implementation, a PCM is given by an Isabelle/HOL type that satisfies a specific typeclass $C$. The set of all values of this type is the carrier set; Typeclass $C$ constrains the presence of the group operation and its PCM axioms.

An Isabelle/HOL type $\alpha$ satisfies typeclass $C$ iff

> there is a set $D : (\alpha \times \alpha)$ set, which defines the domain of the group operation,
> there is a binary operator $(\cdot) : \alpha \to \alpha \to \alpha$ that defines the result of the group operation,
> there is a distinguished constant $\varepsilon : \alpha$ representing the identity element,
> $\forall x.\ x \cdot \varepsilon = x = \varepsilon \cdot x \wedge (x, \varepsilon) \in D \wedge (\varepsilon, x) \in D$
> $\forall (x, y) \in D.\ x \cdot y = y \cdot x \wedge (y, x) \in D$
> $\forall (x, y) \in D, (x \cdot y, z) \in D.\ (y, z) \in D \wedge (x, y \cdot z) \in D \wedge (x \cdot y) \cdot z = x \cdot (y \cdot z)$

In this implementation, SL formulas are represented by a family of Isabelle/HOL types $\{\alpha \text{ set}\}_{\alpha:C}$ indexed by any type $\alpha$ satisfying typeclass $C$. Forcing relation ($\models$) is implemented by the set membership ($\in$). Connectives are implemented by polymorphic types, e.g.,

$(*) : \forall \alpha : C.\ \alpha \text{ set} \to \alpha \text{ set} \to \alpha \text{ set}$     where $D, (\cdot)$ are obtained from
$(\phi * \psi) \triangleq \{w \mid \exists (w_1, w_2) \in D.\ w = w_1 \cdot w_2 \wedge w_1 \in \phi \wedge w_2 \in \psi\}$    the typeclass instance of $\alpha : C$

The reasoning is also parameterized by a type variable $\alpha : C$ that can range over any instance type of typeclass $C$. In this way, the many-sorted logic, and the huge SL model that encompasses any concrete representations and many abstract objects, do not impose much of a burden on both users and reasoner developers.

## C  Examples of Ad-hoc Transformation Rules

As mentioned in §6.3, our system involves ad-hoc transformation rules for eliminating predicate counterparts of logic connectives. The rules for eliminating $(x, y) \mathbin{\S} (T * U)$ are illustrated as follows.

$$\frac{\text{bi-TP}(T_1, U, D_1) \leftarrow (\theta_1, f_1, Z', R_1) \qquad \text{bi-TP}(T_2, Z', D_2) \leftarrow (\theta_2, f_2, Z, R_2)}{\text{bi-TP}((T_1 * T_2), U, D) \leftarrow (\theta_1 \wedge \theta_2, f, Z, R_1 * R_2)} \text{(TP}*_\text{L}\text{)}$$

$$\text{where } f = \lambda((x_1, x_2), w). \text{ let } (w', r_2) = f_2(x_2, w); (y, r_1) = f_1(x_1, w') \text{ in } (y, (r_1, r_2))$$
$$D_1 = \{(x_1, \pi_1(f_2(x_2, z))) \mid ((x_1, x_2), z) \in D\}$$
$$D_2 = \{(x_2, w) \mid ((x_1, x_2), w) \in D\}$$

The rule reduces bi-TP$((T_1 * T_2), U, D)$ to two problems, bi-TP$(T_1, U, D_1)$ and bi-TP$(T_2, Z', D_2)$. To answer bi-TP$((T_1 * T_2), U, D)$, our reasoner first introduces a fresh free variable $f_2$, which occurs in $D_1$. Then, the reasoner tries to solve the sub-problem bi-TP$(T_1, U, D_1)$. Assume the reasoner obtains a solution $(\theta_1, f_1, Z', R_1)$, which means the transformation from $T_1$ to $U$ demands $Z'$ and remains $R_1$. Note that variable $f_2$ can occur in the solution. Next, the reasoner turns to solve bi-TP$(T_2, Z', D_2)$, to extract the demanded $Z'$ from the unused second source $T_2$. If it obtains a solution $(\theta_2, f_2, Z, R_2)$, then initial TP problem bi-TP$((T_1 * T_2), U, D)$ has a solution $(\theta_1 \wedge \theta_2, f, Z, R_1 * R_2)$. Given an assertion $((x_1, x_2), w) \mathbin{\S} (T_1 * T_2) * Z$, this solution first transforms $(x_2, w) \mathbin{\S} (T_2 * Z)$ to $(w', r_2) \mathbin{\S} (Z' * R_2)$ using the solution of the second sub-problem. Then, it transforms $(x_1, w') \mathbin{\S} (T_1 * Z')$ to $(y, r_1) \mathbin{\S} (U * Z')$. Consequently, $(y, (r_1, r_2)) \mathbin{\S} U * (R_1 * R_2)$ is the final end of the transformation indicated by the solution $(\theta_1 \wedge \theta_2, f, Z, R_1 * R_2)$.

The other rule for bi-TP$(T, (U_1 * U_2), D)$ is essentially symmetric to the reduction above. Therefore, we present it as follows without more explanation.

$$\frac{\text{bi-TP}(T, U_1, D_1) \leftarrow (\theta_1, f_1, Z_1, R') \qquad \text{bi-TP}(R', U_2, D_2) \leftarrow (\theta_2, f_2, Z_2, R)}{\text{bi-TP}(T, (U_1 * U_2), D) \leftarrow (\theta_1 \wedge \theta_2, f, Z, R)} \text{(TP}*_\text{R}\text{)}$$

$$\text{where } f = \lambda(x, (w_1, w_2)). \text{ let } (y_1, r') = f_1(x, w_1); (y_2, r) = f_2(r', w_2) \text{ in } ((y_1, y_2), r)$$
$$D_1 = \{(x, w_1) \mid \exists w_2.\ (x, (w_1, w_2)) \in D\}$$
$$D_2 = \{(r', w_2) \mid \exists x\ w_1\ y_1.\ (x, (w_1, w_2)) \in D \wedge (y_1, r') = f_1(x, w_1)\}$$

Additionally, we present the ad-hoc rules for eliminating $x \mathbin{\S} \text{Emp} \triangleq \text{emp} \wedge x = ()$.

$$\frac{\text{Axiom}}{\text{bi-TP}(\text{Emp}, U, D) \leftarrow (D = \{()\},\ \lambda((), w).\ (w, ()),\ U, \text{Emp})} \text{(TP-Emp}_\text{L}\text{)}$$

$$\frac{\text{Axiom}}{\text{bi-TP}(T, \text{Emp}, D) \leftarrow (\text{true},\ \lambda(x, ()).\ ((), x),\ \text{Emp}, U)} \text{(TP-Emp}_\text{R}\text{)}$$

## D  The Detailed Reduction Process of §6.6

Properties hold by Slice:

$$\text{Functor}(\text{Slice}_{[i,j)}, \text{map}, \text{set}) \quad \text{SepHom}(\text{Slice}_{[i,j)}, \text{unzip}, \text{zip}) \quad \text{Dist}(\text{Slice}, \text{split}, \text{cat})$$

$$\text{where map}(f)([l_1, \cdots, l_n]) \triangleq [f(l_1), \cdots, f(l_n)]$$
$$\text{set}([l_1, \cdots, l_n]) \triangleq \{l_1, \cdots, l_n\}$$
$$\text{zip}([a_1, \cdots, a_n], [b_1, \cdots, b_n]) \triangleq [(a_1, b_1), \cdots, (a_n, b_n)]$$
$$\text{unzip}[(a_1, b_1), \cdots, (a_n, b_n)] \triangleq ([a_1, \cdots, a_n], [b_1, \cdots, b_n])$$
$$\text{cat}_{[i,j),[j',k)}(l_1, l_2) \triangleq \text{if } j = j' \text{ then the concatenation of } l_1, l_2 \text{ else undefined.}$$
$$\text{split}_{[i,j),[j',k)}(l) \triangleq \text{if } j = j' \text{ then } ([l_0, \cdots, l_{j-i-1}], [l_{j-i}, \cdots, l_{k-i-1}]) \text{ else undefined.}$$

Given the above properties, SH and SD$_\text{R}$ instantiate the following reasoning rules respectively.

$$\frac{\text{bi-TP}(T, U, D \ggg (d \circ z)) \leftarrow (\theta, f, Z, R)}{\text{bi-TP}(\text{Slice}_{[i,j)}(T), \text{Slice}_{[i,j)}(U), D) \leftarrow (\theta, f', \text{Slice}_{[i,j)}(Z), \text{Slice}_{[i,j)}(R))} \text{ (Slice-SH)}$$
where $f' = \text{unzip} \circ \text{map}(f) \circ \text{zip}$

$$\frac{\text{bi-TP}(\text{Slice}_{[i,k)}(T), F_{[i,k]}(U), h(D)) \leftarrow (\theta, f, Z, R)}{\text{bi-TP}(F_{[i,j)}(T), F_{[i,k)}(U), D) \leftarrow (\theta, g, Z, F_{[k,j)}(T) * R)} \text{ (Slice-SD}_\text{R}\text{) if } k < j$$
where $h = \lambda(x_n, w). \text{let } (x_m, x_\delta) = \text{split}_{[i,k),[k,j)}(x_n)$
$g = \lambda(x_n, w). \text{let } (x_m, x_\delta) = \text{split}_{[i,k),[k,j)}(x_n); (y, r) = f(x_m, w) \text{ in } (y, (x_\delta, r)) \text{ in } (x_m, w)$

Using the above instantiated rules, the bi-TP reduction in §6.6 is detailed as follows.

$$\frac{\text{bi-TP}(\text{Slice}_{[0,N)} \mathbb{Z}, \text{Slice}_{[0,N)} \mathbb{Z}) \leftarrow (\text{true}, \lambda x. x, \text{Emp}, \text{Emp})}{\text{bi-TP}(3) \leftarrow (\text{true}, \lambda(y, r). (y_{[0:N]}, (y_{[N:2N]}, r)), \text{Emp}, \text{Slice}_{[N,2N)} \mathbb{Z} * \text{Emp})} \text{ (Slice-SD}_\text{R}\text{)}$$
by simplification
$$\frac{\text{bi-TP}(3) \leftarrow (\text{true}, \lambda(y, \_). (y_{[0:N]}, y_{[N:2N]}), \text{Emp}, \text{Slice}_{[N,2N)} \mathbb{Z})}{\text{bi-TP}(2) \leftarrow (\text{true},} \text{ (Slice-SH)}$$
$\quad\quad\text{unzip} \circ \text{map}(\lambda(y, \_). (y_{[0:N]}, y_{[N:2N]})) \circ \text{zip},$
$\quad\quad\text{Slice}_{[0,N)} \text{Emp},$
$\quad\quad\text{Slice}_{[0,N)}(\text{Slice}_{[N,2N)} \mathbb{Z}))$
by simplification
$\quad\quad\text{bi-TP}(2) \leftarrow (\text{true},$
$\quad\quad\quad\text{unzip} \circ \text{map}(\lambda y. (y_{[0:N]}, y_{[N:2N]})) \circ \pi_1,$
$\quad\quad\quad\text{Emp},$
$\quad\quad\quad\text{Slice}_{[0,N)}(\text{Slice}_{[N,2N)} \mathbb{Z}))$
Let $\text{cut}_N(y) \triangleq (y_{[0:N]}, y_{[N:2N]})$
$\quad\quad\text{bi-TP}(2) \leftarrow (\text{true},$
$\quad\quad\quad\text{unzip} \circ \text{map}(\text{cut}_N) \circ \pi_1,$
$\quad\quad\quad\text{Emp},$
$\quad\quad\quad\text{Slice}_{[0,N)}(\text{Slice}_{[N,2N)} \mathbb{Z}))$
(Slice-SD$_\text{R}$)
$\quad\text{bi-TP}(1) \leftarrow (\text{true},$
$\quad\quad\lambda(y, \_). \text{let } (y_{AB}, y_{CD}) = \text{cut}_N(y);$
$\quad\quad\quad\quad(y_A, y_B) = \text{unzip}(\text{map}(\text{cut}_N)(y_{AB}))$
$\quad\quad\quad\text{in } (y_A, (y_{CD}, y_B)),$
$\quad\quad\text{Emp},$
$\quad\quad\text{Slice}_{[N,2N)}(\text{Slice}_{[0,N)} \mathbb{Z}) * \text{Slice}_{[0,N)}(\text{Slice}_{[N,2N)} \mathbb{Z})$
$\quad)$

## E  Proofs to some Lemmas

Lemma E.1. *Rule SH is sound.*

Proof. Assume $(\theta, f, Z, R)$ is a solution of bi-TP$(F(T), F(U), D)$. Also assume $\theta$ holds. Starting with $x \mathbin{\$} F(T) * F(Z)$, we can first transform it to $z(x) \mathbin{\$} F(T * Z)$ by property SepHom$(F, s, z)$, then to $(m(f) \circ z)(x) \mathbin{\$} F(U * R)$ by our assumption and property Functor$(F, m, d)$, and finally to $(s \circ m(f) \circ z)(x) \mathbin{\$} F(U) * F(R)$ by property SepHom$(F, s, z)$ again. □

## F  A Systematic Overview for the Templates

Each template defines a reduction aimed at eliminating a predicate operator. The reasoning system composed of rules instantiated from the templates can be understood as recursively performing the following reductions: Given a problem $\mathscr{P}$ that can be either a TP$(T', U', D)$ or a bi-TP$(T', U', D)$,

1: For any predicate operator $F$ having a known property SZero$(F, D')$, if $T'$ matches pattern $F_a(T)$ and ($a$ is a zero element) is provable, apply reduction S0$_\text{L}$; else, if $U'$ matches pattern $F_a(T)$ and ($a$ is a zero element) is provable, apply reduction S0$_\text{R}$.

2: Else, if $(T', U')$ matches pattern $(F(T), F(U))$ for some $F$ satisfying Functor$(F, m, d)$,
3:     if $\mathscr{P}$ is a TP, apply reduction TF;
4:     else, $\mathscr{P}$ must be a bi-TP. Then, check if $F$ has a known property SepHom$(F, s, z)$. If so, apply reduction SH; else, go to step 5.
5: Else, if $(T', U')$ matches pattern $(F_n(T), F_m(U))$ for some known predicate operator $F$, then,
6:     if $(n = m)$ is provable and $F_m$ is a Functor, rewrite $\mathscr{P}$ with $n = m$, and go to step 2;
7:     else, if $F$ has a known property Dist$(F, s, z)$, then,
8:         if $n + \delta = m$ is provable for some $\delta$, apply reduction $\text{SD}_\text{L}$;
9:         else if $n = m + \delta$ is provable for some $\delta$, apply reduction $\text{SD}_\text{R}$;
10:     else, if $F$ has a known property Assoc$(F, g, h)$, then,
11:         if $n \cdot \delta = m$ is provable for some $\delta$, apply reduction $\text{SA}_\text{R}$;
12:         else, if $n = m \cdot \delta$ is provable for some $\delta$, apply reduction $\text{SA}_\text{L}$.
13: Else, if $U'$ matches pattern $F_m(U)$ for some $F$ having a known property SUnit$(F, g, h)$, and if $T'$ does not match $F_n(T)$ for any $n, T$, then apply reduction $\text{S1}_\text{I}$.
14: Else, if $T'$ matches pattern $F_n(T)$ for some $F$ satisfying SUnit$(F, g, h)$, and ($n$ is an identity) is provable, and $U'$ does not match $F_m(U)$ for any $m, T$, then apply reduction $\text{S1}_\text{E}$.

As the reasoning involves arithmetic equations of the ring-like scalar algebra(s), our reasoner has to be parameterized by an automated solver for arithmetics on the algebra(s). In the formalization above, when we state that a formula $P$ is provable, we mean that the formula can be proven by this solver within a time limit.

The reductions for scalar distributivity are incomplete if the scalar addition is not commutative, associative, and cancellative. For bi-TP$(F_n(T), F_m(U), D)$, $\text{SD}_\text{L}$ and $\text{SD}_\text{R}$ only consider the cases when $n = m + \delta$ and $n + \delta = m$ for some $\delta$. However, if the addition is non-commutative (while still assuming associativity and cancellativity), more templates are required to cover the cases of $(\delta' + n + \delta = m)$, $(n + \delta = \delta' + m)$, $(\delta + n = m + \delta')$, and $(n = \delta + m + \delta')$, for some $\delta, \delta'$. For example, continue the Slice example but consider bi-TP$(\text{Slice}_{[i,j]}, \text{Slice}_{[i',k]}, D)$ with $j < k$ and $i < i'$. The bi-TP is irreducible by either $\text{SD}_\text{L}$ or $\text{SD}_\text{R}$, but a template for $n + \delta = \delta' + m$ where we instantiate $n, \delta, \delta', m$ to $[i, j], [j, k], [i, i'], [i', k]$.

Assuming Dist$(F, s, z)$, the templates for the four cases are presented as follows, where we instead use $a, b, c, d, \gamma$ to ranger over scalars.

$$\frac{\text{bi-TP}(F_{d+a}(T), F_{b+c}(U), h(D)) \leftarrow (\theta, f, W, R)}{\begin{array}{l}\text{bi-TP}(F_a(T), F_b(U), D) \leftarrow (\theta, g \circ f \circ h, F_d(T) * W, F_c(U) * R) \\ \text{where } h = (\lambda(x_a, (x_d, w)). (z_{d,a}(x_d, x_a), w)) \\ g = (\lambda(y, r). \text{let } (y_b, y_c) = s_{b,c}(y) \text{ in } (y_b, (y_c, r)))\end{array}}$$

if $a \neq b$ and there are non-zero scalars $\gamma, c, d$ such that $a = \gamma + c \wedge b = d + \gamma$

$$\frac{\text{bi-TP}(F_{a+d}(T), F_{c+b}(U), h(D)) \leftarrow (\theta, f, W, R)}{\begin{array}{l}\text{bi-TP}(F_a(T), F_b(U), D) \leftarrow (\theta, g \circ f \circ h, F_d(T) * W, F_c(U) * R) \\ \text{where } h = (\lambda(x_a, (x_d, w)). (z_{a,d}(x_a, x_d), w)) \\ g = (\lambda(y, r). \text{let } (y_c, y_b) = s_{c,b}(y) \text{ in } (y_b, (y_c, r)))\end{array}}$$

if $a \neq b$ and there are non-zero scalars $\gamma, c, d$ such that $a = c + \gamma \wedge b = \gamma + d$

$$\frac{\text{bi-TP}(F_{a+d}(T), F_{c+b}(U), D) \leftarrow (\theta, f, W, R)}{\begin{array}{l}\text{bi-TP}(F_a(T), F_b(U), D) \leftarrow (\theta, g \circ f, W, F_d(U) * F_c(U) * R) \\ \text{where } g = (\lambda(y, r). \text{let } (y_d, y_{bc}) = s_{d,b+c}(y) \\ (y_b, y_c) = s_{b,c}(y) \\ \text{in } (y_b, (y_d, y_c, r)))\end{array}}$$

if $a \neq b$ and there are non-zero scalars $c, d$ such that $a = d + b + c$

$$\frac{\text{bi-TP}(F_{d+a+c}(T), F_b(U), h(D)) \leftarrow (\theta, f, W, R)}{\text{bi-TP}(F_a(T), F_b(U), D) \leftarrow (\theta, \; f {\circ} h, \; F_d(T){*}F_c(T){*}W, \; R)} \quad \begin{array}{l}\text{if } a \neq b \text{ and there are} \\ \text{non-zero scalars } c, d \\ \text{such that } d + a + c = b\end{array}$$
$$\text{where } h = (\lambda(x_a, (x_d, x_c, w)). \, (z_{d,a+c}(x_d, z_{a,c}(x_a, x_c)), w))$$

## G Overloading & Resolution

Overloading and resolution are essential for state-of-the-art tools like RefinedC to handle low-level programming idioms. In this section, we present an extension of our inference system (primarily the **wp**-transformer and the reduction to bi-EPs) to support overloading and resolution.

### G.1 Introducing Overloading to wp-Transformer

During the backward reasoning process of the **wp**-tranformer (Routine 2), consider a goal $\textbf{wp}_{C(u)}\{v. \, \psi(v)\} \dashv ?$ that infers a pre-condition of program computation $C(u)$. If there are $N$ rules associated with the program $C$, each of which instructs the reasoning process to return a precondition $\phi_i$, the **wp**-transformer returns their disjunction,

$$\textbf{wp}_{C(u)}\{v. \, \psi(v)\} \dashv \phi_1 \vee \cdots \vee \phi_N$$

### G.2 Resolving Overloadings

While state-of-the-art tools typically use pattern matching for overloading resolution, we emphasize the importance of refinement transformations (known as subtypings in refinement-type systems). Therefore, we adopt a semantic proof search strategy based on refinement transformations.

As the extended **wp**-transformer returns a formula involving disjunction $\vee$, we first extend the syntax of goal formulas G to involve disjunction.

$$\text{Goal} \quad \textbf{G} ::= \textbf{S} \mid \textbf{S} * \textbf{G} \mid \textbf{S} \mathrel{-\!\!*} \textbf{G} \mid P \to \textbf{G} \mid \textbf{G} \wedge \textbf{G} \mid \forall \alpha. \, \textbf{G} \mid \exists \alpha. \, \textbf{G} \mid \textbf{G} \vee \textbf{G}$$

To eliminate the disjunction connective, we introduce two rules,

$$\frac{\theta \mid S \vdash G_1}{\theta \mid S \vdash G_1 \vee G_2} \, (\text{G}\vee_\text{L}) \qquad\qquad \frac{\theta \mid S \vdash G_2}{\theta \mid S \vdash G_1 \vee G_2} \, (\text{G}\vee_\text{L})$$

The two rules create a branching point in our reasoning process. When deriving the proof goal $\theta \mid S \vdash G_1 \vee G_2$, our reasoner first attempts the branch applying rule (G$\vee_\text{L}$). The reasoning process continues along this branch, eventually producing a set of bi-EPs and subgoals for eliminating the remains and demands of the bi-EPs. If the subsequent reasoning process successfully solves the bi-EPs and the subgoals, this means that the refinement of state $S$ can transform to $G_1$. Therefore $G_1$ is a valid resolution of the overloading. The reasoning process then proceeds along this branch, discarding the alternative branch for operand $G_2$. Otherwise, if the bi-EPs and subgoals fail to be solved, $G_2$ is then considered as an invalid resolution. The reasoning process backtracks and applies rule (G$\vee_\text{L}$) instead.

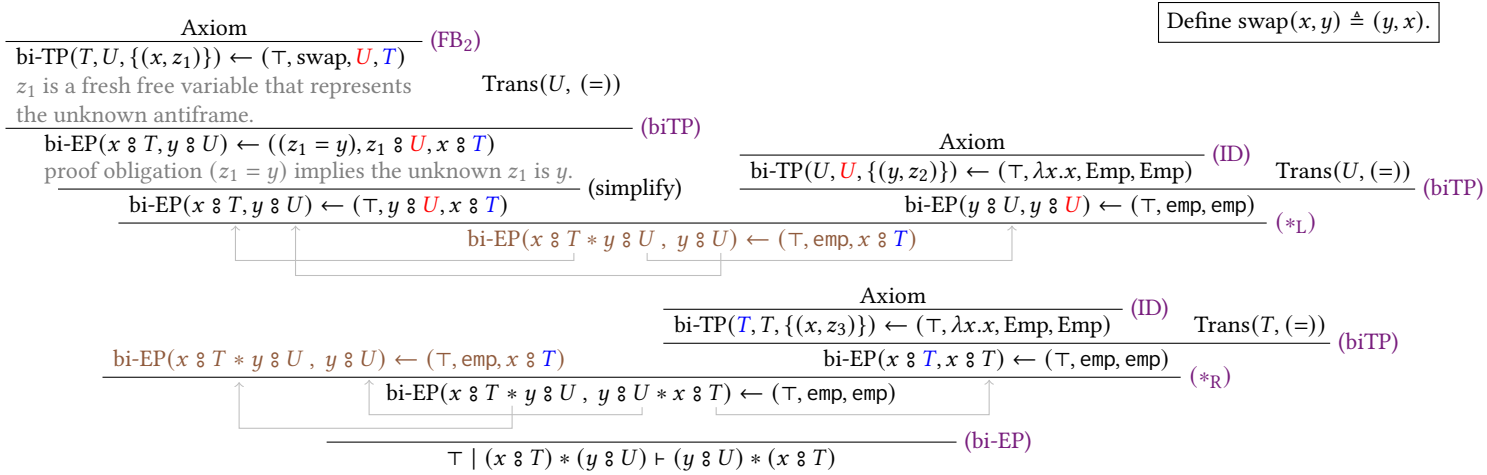

Fig. 9. The derivation of $(x \mathbin{\text{\textsection}} T) * (y \mathbin{\text{\textsection}} U) \longrightarrow (y \mathbin{\text{\textsection}} U) * (x \mathbin{\text{\textsection}} T)$. The tree is too large, so we split it into two parts linked by brown. Blue color denotes a frame and red denotes an antiframe. Gray arrows help readers to trace the predicates.

## H Example: Deriving $(x \mathbin{\text{\textsection}} T) * (y \mathbin{\text{\textsection}} U) \longrightarrow (y \mathbin{\text{\textsection}} U) * (x \mathbin{\text{\textsection}} T)$

Instead of permuting predicates on the left-hand sides, nor matching the left-hand side items with the right-hand side, our reasoner decomposes $(x \mathbin{\text{\textsection}} T) * (y \mathbin{\text{\textsection}} U) \longrightarrow (y \mathbin{\text{\textsection}} U) * (x \mathbin{\text{\textsection}} T)$ into three bi-TPs as illustrated in Fig. 9. First, by $(*_R)$, $(*_L)$, we extract the first target $y \mathbin{\text{\textsection}} U$ from the first source item $x \mathbin{\text{\textsection}} T$, written bi-EP$(x \mathbin{\text{\textsection}} T, y \mathbin{\text{\textsection}} U)$. It reduces to bi-TP$(T, U, \{x\})$ by (biTP). As no bi-TP rule is further applicable, fallback (FB$_2$) is called, leaving the entire $x \mathbin{\text{\textsection}} T$ as the remaining source and the entire $y \mathbin{\text{\textsection}} U$ as the unfulfilled target. Next, as stated in rule $(*_R)$ and $(*_L)$, the reasoning process turns to extract the unfulfilled target from the second source item and to extract the second target from the remaining source, i.e., bi-EP$(y \mathbin{\text{\textsection}} U, y \mathbin{\text{\textsection}} U)$ and bi-EP$(x \mathbin{\text{\textsection}} T, x \mathbin{\text{\textsection}} T)$, both of which trivially succeed.

## I The Module-like Algebra of Linked List Segment

This section means to show that the data structure of Linked List segment also satisfies the (relaxed) model of modules over rings.

- Let $l \,\S\, \text{Lseg}_a \xrightarrow{n} {}_b$ represent a linked list Segment having head address $a$, tail address $b$ and length $n$. The abstraction of this segment is denoted by a logical list $l$.
- A scalar is a labelled arrow $\text{Lseg}_a \xrightarrow{n} {}_b$ from the head node to the tail node, with the length as its label. The scalar addition is the arrow concatenation (with adding the labelled length), and a zero scalar is a 0-length loop.

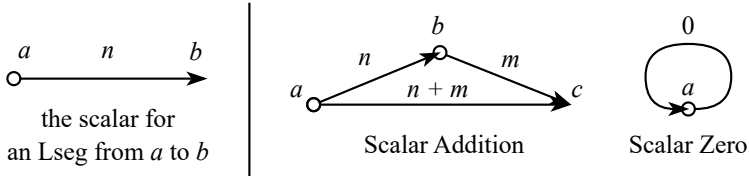

the scalar for an Lseg from $a$ to $b$ | Scalar Addition | Scalar Zero

——— Laws ———

Distributivity
$$\left( \begin{array}{l} \text{Lseg}_a \xrightarrow{n} {}_b * \text{Lseg}_b \xrightarrow{m} {}_c \xrightarrow{\text{cat}} \text{Lseg}_a \xrightarrow{n+m} {}_c \\ \exists b.\ \text{Lseg}_a \xrightarrow{n} {}_b * \text{Lseg}_b \xrightarrow{m} {}_c \xleftarrow{\text{cut}} \text{Lseg}_a \xrightarrow{n+m} {}_c \end{array} \right)$$

Identity
$$[x] \,\S\, \text{Lseg}_a \xrightarrow{1} {}_b \iff (\textit{data}\colon x,\ \textit{nxt}\colon b) \,\S\, \text{Node}_a$$

Zero
$$\text{Lseg}_a \xrightarrow{0} {}_a \iff \text{Empty}$$

The red-marked existential quantification might suggest the demand of an extension to the system presented in the paper.